\documentclass[twocolumn,prb]{revtex4}
\usepackage{amsmath}
\usepackage{amssymb}
\usepackage{esint}

\makeatletter
\@ifundefined{textcolor}{}
{%
 \definecolor{BLACK}{gray}{0}
 \definecolor{WHITE}{gray}{1}
 \definecolor{RED}{rgb}{1,0,0}
 \definecolor{GREEN}{rgb}{0,1,0}
 \definecolor{BLUE}{rgb}{0,0,1}
 \definecolor{CYAN}{cmyk}{1,0,0,0}
 \definecolor{MAGENTA}{cmyk}{0,1,0,0}
 \definecolor{YELLOW}{cmyk}{0,0,1,0}
 }

\@ifundefined{definecolor}
 {\@ifundefined{definecolor}
 {\@ifundefined{definecolor}
 {\usepackage{color}}{}
}{}
}{}

\newcommand{\Ket}[1]{\vert  #1  \rangle}

\newcommand{\MatEl}[3]{\langle \, #1 \,\vert\,#2\,\vert\,#3\,\rangle}
\newcommand{\Amp}[2]{\langle \, #1\, \vert\,  #2 \, \rangle}

\newcommand{\Avg}[1]{\langle  #1  \rangle}

\renewcommand{\phi}{\varphi}
\renewcommand{\epsilon}{\varepsilon}
\renewcommand{\vec}[1]{{\bf #1}}

\begin{document}

\title{Semi-classical model for the dephasing of a two-electron spin qubit
coupled to a coherently evolving nuclear spin bath}

\author{Izhar Neder}
\affiliation{Department of Physics, Harvard University, Cambridge, MA 02138, USA}

\author{Mark S. Rudner}
\affiliation{Department of Physics, Harvard University, Cambridge, MA 02138, USA}

\author{Hendrik Bluhm}
\altaffiliation[Present address: ]{2nd Institute of Physics C, RWTH Aachen University, 52074 Aachen, Germany}
\affiliation{Department of Physics, Harvard University, Cambridge, MA 02138, USA}

\author{Sandra Foletti}
\affiliation{Department of Physics, Harvard University, Cambridge, MA 02138, USA}

\author{Bertrand I. Halperin}
\affiliation{Department of Physics, Harvard University, Cambridge, MA 02138, USA}

\author{Amir Yacoby}
\affiliation{Department of Physics, Harvard University, Cambridge, MA 02138, USA}

\begin{abstract}
We study electron spin decoherence in a two-electron double quantum
dot due to the hyperfine interaction, under spin-echo conditions as
studied in recent experiments. We develop a semi-classical model for
the interaction between the electron and nuclear spins, in which the
time-dependent Overhauser fields induced by the nuclear spins are
treated as classical vector variables. Comparison of the model with
experimentally-obtained echo signals allows us to quantify the contributions
of various processes such as coherent Larmor precession and spin diffusion
to the nuclear spin evolution. 
\end{abstract}
\maketitle

\section{Introduction}

In recent years, electron spin qubits in solid-state quantum dots
have emerged as promising candidates for the implementation of quantum
information processing\cite{Petta2005,Koppens2008,Greilich2006,Foletti2009,Nowack2007,Barthel2009,Loss1998,Taylor2005}.
The confined electrons in these devices can be precisely manipulated
using microwave frequency electromagnetic fields and/or nanosecond-scale
pulses of nearby electrostatic gates, while maintaining spin coherence
over much longer times. The main source of decoherence in such qubits
is the hyperfine interaction between the electron spins and the nuclear
spins of the host lattice \cite{Taylor2005,Taylor2007,HansonRMP,Merkulov2002,Erlingsson2002,Johnson2005,Sousa2003,Koppens2005}.
Through this interaction, the nuclear spin bath produces a fluctuating
effective Zeeman field on the electron spins. However, the timescale
for evolution of this so-called {}``Overhauser field'' is typically
much longer than that required for manipulation of a single qubit.
Therefore, dynamical decoupling techniques\cite{Hahn1950,Meiboom1958}
based on fast control of the qubit can be employed to partially eliminate
decoherence due to the interaction with the nuclear spins. Recent
experiments confirm that such techniques can be used to extend qubit
coherence times by a few orders of magnitude, up to approximately
$200\ \mu$s \cite{Petta2005,ourselves}.

Usually, interactions between a single electron spin and many weakly
interacting spins (as described by the {}``central spin problem''\cite{Gaudin1976})
can lead to complicated evolution of the joint quantum system\cite{Khaetskii2002,Erlingsson2004,Yao2006,Yao2007,Deng2006,Coish2004,Witzel2005,Shenvi2005,Al-Hassanieh2006,Coish2008,Chen:semiclass, Rashba2008}.  Moreover, the nuclear spin bath may maintain coherence over a long
time, which may in general result in coherent back-action on the electron
spin. Indeed, using the Keldysh technique and a non-trivial re-summation
of diagrams, the authors of Refs.~\onlinecite{ Cywinski2009,Cywinski2009B}
analyzed this problem, and predicted periodic collapses and revivals
of the electron spin coherence over a specific range of external magnetic
field strengths. This phenomenon was subsequently observed in two-electron
spin echo measurements\cite{ourselves}.

Despite the apparent complexity of the system, there are several reasons
why one might expect to find a more intuitive semi-classical description
of the electron spin dynamics. First, due to the small nuclear Zeeman
energy, the initial state of the nuclear spin bath is well described
by an infinite temperature (completely random) state. Second,
 the state of the nuclear spin bath is not measured in the
experiment, and the experimental outcome is an average over many runs. We note that the semi-classical approximation of the nuclear spin system has been used to describe a variety of other interesting phenomena in quantum dots  \cite{Erlingsson2004,Al-Hassanieh2006,Chen:semiclass, Rashba2008}.

The aim of this paper is to demonstrate that a wide array
of complex dynamical phenomena in two-electron spin echo measurements,
such as those of Ref.~\onlinecite{ourselves}, can be understood
within the context of a semi-classical treatment of the nuclear spin
bath. We first show that, within a simple model which neglects the
effects of the Knight shift and the dipolar interaction between nuclear spins,
the semi-classical treatment reproduces the expressions for the spin echo signal obtained in Refs.~\onlinecite{ Cywinski2009,Cywinski2009B}, where a summation of diagrams in a perturbative quantum mechanical treatment was used.
We then present a more detailed microscopic model, 
which incorporates the Knight shift and the nuclear dipole-dipole interactions, as well as inhomogeneous hyperfine and Zeeman couplings. 
The semi-classical treatment for this model was sketched in the supplementary material
of Ref.~\onlinecite{ourselves}. 
Here we provide a systematic discussion of this semi-classical treatment, which relies on a low-order
expansion of the qubit evolution in the inverse of the number of nuclear spins and in the hyperfine coupling. 
Using this approach, we identify
the relevant physical processes which govern nuclear spin evolution
and electron spin coherence, as measured by the spin-echo signal.
The excellent agreement between the model and the experimental data
that was shown in Ref.~\onlinecite{ourselves} lends additional
justification to the approximations underlying the semi-classical
approach.

The paper is organized as follows. In Sec.~\ref{sec:genral-remarks}
we describe the two-electron-spin qubit system and outline the main
physical processes which govern the qubit dynamics. In Sec.~\ref{spinecho}
we review the Hahn spin-echo sequence. In Sec.~\ref{sec:Hamiltonian}
we describe the Hamiltonian of the two-electron-spin system and the
nuclear spin bath. We then derive a simpler effective Hamiltonian
through a perturbative treatment of the Overhauser field. In Sec.~\ref{sec:simplest_model}
we show that a semi-classical model based on this Hamiltonian reproduces
the collapse and revival phenomenon. A more complete semi-classical
approach is derived and justified in Sec.~\ref{sec:SC_derivation}.
There we include the effect of the dipolar coupling between nuclear
spins and the back-action of the Knight field on the nuclear spin
precession. Conclusions and discussion are presented in Sec.~\ref{sec:Discussion}.

\section{The two-electron-spin qubit}

\label{sec:genral-remarks}

We consider a qubit consisting of two electrons in a double quantum
dot, in the regime where the two electrons are well separated with
one electron occupying each dot. A uniform in-plane magnetic field
$\mathbf{B}_{{\rm ext}}=B_{{\rm ext}}\hat{\vec{z}}$ is applied 
along the $z$-axis, with $B_{{\rm ext}}>0$. 
The qubit Hilbert space is spanned by the singlet spin state, $\left|S\right\rangle =\frac{1}{\sqrt{2}}\left(\left|\uparrow\downarrow\right\rangle -\left|\downarrow\uparrow\right\rangle \right)$,
and the triplet state with zero net spin projection on the $z$-axis,
$\left|T_{0}\right\rangle =\frac{1}{\sqrt{2}}\left(\left|\uparrow\downarrow\right\rangle +\left|\downarrow\uparrow\right\rangle \right)$.
In this notation, the two arrows represent the orientations of the
electron spins in each of the two dots, measured relative to the external
field direction, $\hat{\vec{z}}$. 
Due to the large Zeeman splitting, this qubit subspace is energetically
isolated from the two other two-electron triplet spin states $\left|T_{+}\right\rangle =\left|\uparrow\uparrow\right\rangle $
and $\left|T_{-}\right\rangle =\left|\downarrow\downarrow\right\rangle $,
in which both electron spins point parallel or anti-parallel to the
direction of the magnetic field. 

The energy splitting between the two qubit states $\left|S\right\rangle $
and $\left|T_{0}\right\rangle $ can be controlled on a fast timescale
by rapidly tuning nearby electrostatic gates which modify the confining potential for the quantum dot, and hence control the shape of the two-electron wave function.
For tunings where the two electrons are held far apart in the ground state, i.e. where the electrons are separated into different dots,
the states $\Ket{S}$ and $\Ket{T_{0}}$ are degenerate.
However, when the potential is tuned to favor partial double occupation of one dot, 
the difference in orbital symmetry between $\Ket{S}$ and $\Ket{T_{0}}$ leads to an exchange energy splitting $J$ between them.

In materials such as the commonly employed III-V compounds, the confined
electron spins interact with a background of nuclear spins in the
host lattice. This interaction is produced by the hyperfine coupling
$H_{{\rm HF}}=\sum_{d,n}A_{d,n}\vec{I}_{n}\cdot\vec{S}_{d}$. Here
the index $n$ labels all nuclear spins, described by the operators
$\{\vec{I}_{n}\}$, $d=L,R$ labels the electron spins in the left
and right dots, described by the operators $\{\vec{S}_{d}\}$, and
the coupling constants $\{A_{d,n}\}$ depend on the local electron
spin density, as will be described in detail below. Defining the nuclear
(Overhauser) field operator in dot $d$ as $g^{*}\mu_{B}\vec{B}_{{\rm nuc},d}\equiv\sum_{n}A_{d,n}\vec{I}_{n}$,
we write the ``effective'' electron spin Zeeman Hamiltonian as
\begin{equation}
H_{el}=g^{*}\mu_{B}\left(\mathbf{S}_{L}\cdot\mathbf{B}_{{\rm tot},L}+\mathbf{S}_{R}\cdot\mathbf{B}_{{\rm tot},R}\right),\label{eq:Hel}\end{equation}
 with \begin{equation}
\mathbf{B}_{{\rm tot},d}=\mathbf{B}_{{\rm ext}}+\mathbf{B}_{{\rm nuc},d}.\label{eq:Btot}\end{equation}
 Here $g^{*}\approx-0.4$ is the electron effective g-factor in GaAs
and $\mu_{B}$ is the Bohr magneton. Equations (\ref{eq:Hel}) and
(\ref{eq:Btot}) describe the Zeeman coupling in a system of two isolated
electrons, where each electron is subjected to an effective field
which is the vector sum of a uniform static external magnetic field
$\mathbf{B}_{{\rm ext}}$, and a local, operator-valued, Overhauser
field $\mathbf{B}_{{\rm nuc},d}$.

We can gain extremely useful intuition about electron spin dynamics
in this system by treating the operator-valued Overhauser fields $\vec{B}_{{\rm nuc},L}$
and $\vec{B}_{{\rm nuc},R}$ as classical (time-dependent)
vector variables. 
In typical GaAs dots, the Overhauser field is produced 
by a large number of nuclear spins in each dot, $N_{d}\approx10^{6}$.
When all the nuclear spins are polarized, the resulting effective Overhauser
field has a magnitude $|\mathbf{B}_{{\rm nuc}, d}| \sim 5$ T. 
However, under experimental conditions, where thermal fluctuations randomize the directions of all nuclear spins,
the typical values of $\left|\mathbf{B}_{{\rm nuc},d}\right|$ are reduced by a factor $\sqrt{N_{d}}$, and are of order $1$ mT.
For strong enough external fields, $B_{{\rm ext}}\gg\left|\mathbf{B}_{{\rm nuc},d}\right|$,
the net fields $\mathbf{B}_{{\rm tot},L}$ and $\mathbf{B}_{{\rm tot},R}$
are nearly parallel to $\vec{B}_{{\rm ext}}$. Under these conditions,
the two-dimensional qubit subspace is only slightly perturbed by the
misalignment of local fields, and remains energetically isolated from
the other two-electron spin states. To leading order in $\frac{\left|\mathbf{B}_{{\rm nuc},d}\right|}{B_{{\rm ext}}}$,
the effect of the nuclear fields is simply to induce a Zeeman splitting
between $\left|\downarrow\uparrow\right\rangle $ and $\left|\uparrow\downarrow\right\rangle $,
proportional to the difference in $z$-projections of the effective
fields in the two dots, $\Delta B_{{\rm nuc}}^{z}=B_{{\rm nuc},L}^{z}-B_{{\rm nuc},R}^{z}$.

If we define a  Bloch sphere for the qubit, whose poles on the z axis are the states $\left|\downarrow\uparrow\right\rangle$ and $\left|\uparrow\downarrow\right\rangle$, then the field  $\Delta B_{{\rm nuc}}^{z}$ points along the z axis. The states $\Ket{S}$ and $\Ket{T_{0}}$ lie on the x axis of this Bloch sphere.
The splitting induced by $\Delta B_{{\rm nuc}}^{z}$ leads to oscillations
between the qubit states $\Ket{S}$ and $\Ket{T_{0}}$, with a frequency
proportional to $|\Delta B_{{\rm nuc}}^{z}|$. Such oscillations are
polluted, however, by two sources of randomness. First, due to the
fact that the nuclear state is random for an equilibrium nuclear spin
bath, the magnitude of the initial nuclear field $|\Delta B_{{\rm nuc}}^{z}|$,
and hence the initial frequency of oscillations, is unknown. Furthermore,
due to internal dynamics of the bath itself, the nuclear fields $\vec{B}_{{\rm nuc},d}(t)$
evolve in time. The resulting ``spectral diffusion'' of the qubit
oscillation frequency leads to dephasing of the qubit oscillations\cite{Klauder1962}.
Using an electron-spin-echo pulse, as explained below in Sec.~\ref{spinecho},
dephasing due to the unknown \textit{mean value} of $\Delta B_{{\rm nuc}}^{z}(t)$
over some interval can be reversed. However, decoherence due to fluctuations
of $\Delta B_{{\rm nuc}}^{z}(t)$ on a timescale comparable to or
shorter than the period between echo pulses in general cannot be eliminated
in this way.

For experiments involving weaker external magnetic fields and long
enough evolution times, it is necessary to go beyond the leading order
in $\frac{\left|\mathbf{B}_{{\rm nuc},d}\right|}{B_{{\rm ext}}}$.
Here we find that the transverse components of the Overhauser field,
$B_{{\rm nuc},d}^{x,y}(t)$, crucially affect the qubit evolution
in two primary ways. First, these transverse field components contribute
to qubit decoherence by causing leakage of the electron spin state
into the {}``non-qubit'' subspace spanned by the states $\left|\downarrow\downarrow\right\rangle $
and $\left|\uparrow\uparrow\right\rangle $. Second, the magnitude
of the transverse part of the Overhauser field introduces a correction
to the frequency of the $\Ket{S}-\Ket{T_{0}}$ oscillations described
above. Note that while in general a spin-echo pulse cannot reverse
dephasing due to time-dependent fluctuations in the local fields,
a partial or full recovery is possible if these fields vary \textit{periodically}
in time. Such a periodic time dependence, produced by the relative
Larmor precession of different nuclear species, is the origin of the
collapse and revival phenomenon.

An added complication in this moderate field regime arises from the
fact that, when we treat the Overhauser fields more properly as quantum
mechanical operators, $B_{{\rm nuc},d}^{x}(t)$, $B_{{\rm nuc},d}^{y}(t)$
and $B_{{\rm nuc},d}^{z}(t)$ do not commute. Consequently, at this
order, the semi-classical approach to electron spin dephasing requires
more justification. As we discuss below, such an approach is valid
when the number of nuclear spins is large, and when the interaction
between the electronic and nuclear spins is weak.

\section{The Spin-Echo Sequence\label{spinecho}}

With the above-described picture in mind, below we focus on the Hahn
echo experiment in GaAs double quantum dots (see e.g.~Ref.~\onlinecite{ourselves}),
where each electron spin interacts weakly with $N\approx10^{6}$ nuclear
spins. In such experiments, the two-electron state in the double quantum
dot is initialized to the singlet ground state at large potential
detuning, where both electrons reside in the right dot, $\left|(0,2)S\right\rangle $.
Here the numbers in parentheses indicate the electron occupation numbers
in the left and right dots, respectively, and the letter $S$ indicates
the two-electron (singlet) spin state. By rapidly tuning the potentials
of nearby electrostatic gates, one of the electrons is transferred
to the left dot within a time scale of approximately $1$ ns. After
this operation, the two separated electron spins evolve freely for
a time $\frac{\tau}{2}$ under the influence of the net local fields
$\vec{B}_{{\rm tot},d}$, given in Eq.~(\ref{eq:Btot}). 
The gate
potentials are then rapidly tuned to bring the electrons closer together.
Here a substantial exchange energy splitting between $\left|S\right\rangle $
and $\left|T_{0}\right\rangle $ is maintained for a time corresponding
to a {}``$\pi$-phase'' duration, which effectively leads to the
the swapping of the states $\left|\downarrow\uparrow\right\rangle \leftrightarrow\left|\uparrow\downarrow\right\rangle $.
Then the gate voltages are rapidly tuned to separate the electrons,
and the system is allowed to evolve over another interval of length
$\frac{\tau}{2}$.

At the end of the cycle, a spin readout procedure is performed. The
gates are rapidly tuned to a large positive potential bias, where
the singlet ground state takes the ``(0,2)'' orbital configuration,
while the orbital part of the triplet state remains of the ``(1,1)''
type due to Pauli exclusion. The charge configuration is then measured
via a nearby charge sensitive detector. Due to the correlation between
the orbital and spin degrees of freedom, the final spin state of the
two electrons can be inferred from this charge measurement. Ignoring
any imperfections of the measurement itself, we assume that the average
of the charge detector signal taken over many runs depends linearly
on the singlet return probability.

The detector signal is averaged over a timescale which is long compared
with all correlation times of the nuclear spin bath. Therefore we
equate the averaged singlet return probability for evolution duration
$\tau$, $P_{S}(\tau)$, with the average of single-run singlet return
probabilities, taken over the equilibrium ensemble of initial nuclear
spin configurations. Note that because the qubit is initialized in
the singlet state, $P_{S}(\tau)$ approaches 1 for very short evolution
times $\tau$. For very long times, when coherence is lost and the
qubit tends to an equal-probability classical mixture of the states
$\Ket{S}$ and $\Ket{T_{0}}$, $P_{S}(\tau)$ approaches 1/2. Therefore,
by convention, we define the echo signal as $2P_{S}(\tau)-1$, which
takes the value 1 for $P_{S}(\tau)=1$ and 0 for $P_{S}(\tau)=1/2$.
In this sense, the echo signal is used as a measure of electron spin
coherence.

In the recent experiment of Ref.~\onlinecite{ourselves}, the echo
signal was observed to decay monotonically on a timescale of approximately
$30$ $\mu$s in high external magnetic fields above $300$~mT. At
intermediate magnetic fields ($120-300$~mT) additional small fast
oscillations were observed. At lower magnetic fields ($50-120$~mT)
these oscillations evolved into a complete collapse of the echo signal
at $\tau\approx1\,\mu$s, followed by a pattern of revivals and collapses
on a timescale $\tau\approx10\,\mu$s. The collapse and revival pattern
was attenuated by a decaying envelope over a timescale of $\tau\approx30\,\mu$s.

Our theoretical analysis of the echo experiments rests on a separation
of timescales in the dynamics of the system. First, if one omits the
$\pi$-pulse from the experimental protocol, $P_{S}(\tau)$ for the
{}``free induction decay'' decays to the value of $1/2$ on the
timescale $T_{2}^{*}\approx10$ ns\cite{Petta2005,Taylor2005}. This
decay results from the uncertainty in the $z$ component of the Overhauser
field, $\Delta B_{{\rm nuc}}^{z}$, which varies from run to run.
In contrast, in a spin-echo measurement, the influence of a random,
static Overhauser field $\Delta B_{{\rm nuc}}^{z}$ on the final electron
spin state is eliminated by the combination of the $\pi$ pulse and
the two equal-length free evolution periods. If the Overhauser field
were truly static, the electron spins would return to the state $\left|S\right\rangle $
at the end of the evolution. In a perfect measurement, for such a
static Overhauser field, one would then obtain $P_{S}(\tau)=1$, or
an echo signal of value 1. Due to the time dependence of $\mathbf{B}_{{\rm nuc},d}(t)$,
however, the echo signal typically decays to zero on a timescale of tens of
microseconds.

During the free evolution time while the electrons are separated,
the system exhibits oscillations between $\Ket{S}$ and $\Ket{T_{0}}$.
Our crucial finding is that, 
for $\frac{|\vec{B}_{{\rm nuc},d}|}{B_{{\rm ext}}} \ll 1$, the
oscillations are well described in terms of the net accumulated phase
determined by difference of magnitudes of the \textit{total} effective
fields on the two dots (we take $\hbar=1$): \begin{equation}
\Delta\Phi(t)=g^{*}\mu_{B}\int_{0}^{t}\Big[\left|\mathbf{B}_{{\rm tot},L}(t')\right|-\left|\mathbf{B}_{{\rm tot},R}(t')\right|\Big]dt'.\label{eq: phi_Btot}\end{equation}
 The magnitude of the total field is given by \begin{equation}
\left|\vec{B}_{{\rm tot},d}(t)\right|=\sqrt{\left(B_{{\rm ext}}+B_{{\rm nuc},d}^{z}\right)^{2}+|\vec{B}_{{\rm nuc},d}^{\perp}(t)|^{2}},\label{eq:Btotmag}\end{equation}
 where $|\vec{B}_{{\rm nuc},d}^{\perp}(t)|$ is the magnitude of the
component of the Overhauser field in dot $d$ which is perpendicular
to the external magnetic field. For $B_{{\rm ext}}$ not too large,
$\left|\vec{B}_{{\rm tot},d}(t)\right|$ includes a significant contribution
from $|\vec{B}_{{\rm nuc},d}^{\perp}(t)|$. Note that the time dependence
of $\vec{B}_{{\rm nuc},d}^{\perp}(t)$ is dominated by the relative
Larmor precession of the three nuclear spins species, $^{69}$Ga,
$^{71}$Ga and $^{75}$As. Such relative precession leads to a time-dependent
modulation of $|\vec{B}_{{\rm tot},d}|$ and causes a reduction of
$P_{S}(\tau)$ on a timescale of microseconds. In addition, random
fluctuations of $B_{{\rm nuc},d}^{z}(t)$, which arise due to interactions
between nuclear spins, lead to a reduction of $P_{S}(\tau)$ on the
same timescale. To account for all the above-mentioned effects, which
were observed in Ref.~\onlinecite{ourselves}, we thus include
a combination of deterministic (Larmor precession) and stochastic
processes in the evolution of the Overhauser fields in our semi-classical
approach.

\section{The qubit Hamiltonian}

\label{sec:Hamiltonian}

We begin our quantitative investigation by constructing the quantum
Hamiltonian that describes the spin echo experiment. Although the
experiment consists of several temporal stages, we focus on the free
evolution periods, during which the Overhauser fields exert their
main influence on the qubit evolution. We consider all other stages
of the experiment, i.e. the singlet state initialization at the first
stage, the pi pulse, and the measurement of the final state, to be perfect.

During the free evolution period while the electrons are well-separated,
the qubit state evolves according to the effective Zeeman Hamiltonian
\begin{equation}
H_{el}=g^{*}\mu_{B}\left(\mathbf{S}_{L}\cdot\mathbf{B}_{{\rm tot},L}+\mathbf{S}_{R}\cdot\mathbf{B}_{{\rm tot},R}\right),\label{eq: H}\end{equation}
 presented above in Eq.~(\ref{eq:Hel}). The total effective field
in dot $d$, $\vec{B}_{{\rm tot},d}$, is formed by the vector sum
of the uniform external field $\vec{B}_{{\rm ext}}$, and the Overhauser
field

\begin{equation}
g^{*}\mu_{B}\vec{B}_{{\rm nuc},d}=\sum_{n}A_{d,n}\mathbf{I}_{n},\quad A_{d,n}=\mathcal{A}_{\alpha(n)}|\psi_{d,n}|^{2},\label{eq:Bnuc}\end{equation}

where $n$ is a label which indexes all of the nuclei. The parameter
$\mathcal{A}_{\alpha(n)}$ is the microscopic hyperfine coupling for
nuclear spin species $\alpha(n)$, while the factor $|\psi_{d,n}|^{2}$
weights the hyperfine coupling to nuclear spin $n$ in dot $d$ by
the local electron density, and satisfies  the normalization condition
$\sum_{n}|\psi_{d,n}|^{2}=n_c$, where $n_c=2$ is the number of nuclei per unit cell of the GaAs lattice. In regions where the electron density
is substantial, the coupling to nuclei goes as $|\psi_{d,n}|^{2}\sim\frac{1}{N_{d}}$,
where we define $N_{d}\equiv n_c^2/\sum_{n}|\psi_{d,n}|^{4}$ as the effective
number of nuclei in dot $d$. For typical GaAs quantum dots, $N_{d}\approx(1-4)\cdot10^{6}$.
The index $\alpha=\{1,2,3\}$ runs over the three nuclear species
$^{69}$Ga, $^{71}$Ga, and $^{75}$As.

To describe the evolution of the nuclear spin bath, we employ a Hamiltonian
which includes Zeeman terms with a species- and site-dependent Larmor
frequency for each nuclear spin, and the dipolar coupling between
all pairs of nuclei: \begin{equation}
H_{{\rm nuc}}=\sum_{n}\omega_{n}I_{n}^{z}+\sum_{n,n'}D_{n,n'}^{ij}I_{n}^{i}I_{n'}^{j},\label{eq:Hnuc}\end{equation}
 where $n$ and $n'$ run over all nuclei, and $i$ and $j$ label
the Cartesian components of the nuclear spin operators.

We wish to identify the main sources of decoherence for the two electron
spin qubit which arise from the combined evolution under the Hamiltonian
in Eqs.~(\ref{eq:Hel}) and (\ref{eq:Hnuc}). Naturally, we shall
use $1/N_{d}$ as a small parameter. In addition, we note that the
lowest external field used in the experiment in Ref.~\onlinecite{ourselves}
($50$ mT) was more than an order of magnitude bigger than the typical
magnitude of the Overhauser field. Therefore we will proceed to study
decoherence effects 
as an expansion in 
$\frac{\left|\vec{B}_{{\rm nuc},d}\right|}{B_{{\rm ext}}}$,
and to leading order in $1/N_{d}$.

First, note that if one replaces the operators $\mathbf{B}_{{\rm nuc},L(R)}$
in Eq.~(\ref{eq:Btot}) by classical vectors with magnitudes much
smaller than $B_{{\rm ext}}$, then, similar to the case when the
Overhauser field was absent, the system described by Hamiltonian (\ref{eq:Hel})
possesses a two-dimensional subspace which is energetically well-separated
from the remaining two levels. This new qubit subspace is spanned
by the states $\left|\uparrow_{\hat{n}_{L}}\right\rangle \otimes\left|\downarrow_{\hat{n}_{R}}\right\rangle $
and $\left|\downarrow_{\hat{n}_{L}}\right\rangle \otimes\left|\uparrow_{\hat{n}_{R}}\right\rangle $,
with eigenvalues $\pm\frac{1}{2}\mu_{B}g^{*}\left(\left|\mathbf{B}_{{\rm tot},L}\right|-\left|\mathbf{B}_{{\rm tot},R}\right|\right)$.
Here the up and down arrows indicate the projections of the electron
spins onto the quantization axes $\hat{n}_{L}$ and $\hat{n}_{R}$
parallel to the total fields $\vec{B}_{{\rm tot},L(R)}$ in each dot.

Deviations of the directions of $\hat{n}_{L}$ and $\hat{n}_{R}$
from the $z$-axis arise from the Overhauser field components perpendicular
to the applied field, $\vec{B}_{{\rm nuc},d}^{\perp}$. Due to evolution
of the nuclear spin bath (primarily due to Larmor precession of the
nuclear spins), the fields $\vec{B}_{{\rm nuc},d}^{\perp}(t)$, and
hence $\hat{n}_{L}$ and $\hat{n}_{R}$, are slowly modulated in time.
However, because the frequencies of such modulations are typically
two orders of magnitude smaller than the value of the energy gap between
the two instantaneous eigenstates, $\Delta E=\mu_{B}g^{*}\left(\left|\mathbf{B}_{{\rm tot},L}\right|-\left|\mathbf{B}_{{\rm tot},R}\right|\right)$,
we assume that each electron spin adiabatically follows its local,
slowly varying, quantization axis $\hat{n}_{L}$ or $\hat{n}_{R}$.
For small $\frac{\left|\vec{B}_{{\rm nuc},d}\right|}{B_{{\rm ext}}}$,
the main effect of the nuclear field is thus to modulate the magnitude
of the total field $\left|\vec{B}_{{\rm tot},d}(t)\right|$, Eq.~(\ref{eq:Btotmag}),
and hence to modify the dynamical phase $\int\Delta E(t)dt$ accumulated
between the two eigenstates over the free evolution period. We therefore
ignore changes in the directions of the quantization axes in each
dot, and describe the evolution of the system by using the effective
Hamiltonian \begin{equation}
H_{el,z}=g^{*}\mu_{B}\left(S_{L}^{z}\left|B_{{\rm tot},L}\right|+S_{R}^{z}\left|B_{{\rm tot},R}\right|\right).\label{eq:H_z}\end{equation}
 For given classical values of $\mathbf{B}_{{\rm tot},L}$ and $\mathbf{B}_{{\rm tot},R}$,
this Hamiltonian preserves the instantaneous eigenvalues of the original
Hamiltonian $H_{el}$, Eq.~(\ref{eq:Hel}). Expanding $\left|\vec{B}_{{\rm tot},d}\right|$
in Eq.~(\ref{eq:Btotmag}) in the small parameter $\frac{\left|\vec{B}_{{\rm nuc},d}\right|}{|B_{{\rm ext}}|}$,
the Hamiltonian $H_{el,z}$ becomes \begin{eqnarray}
H_{el,z}\approx g^{*}\mu_{B}\sum_{d=L,R}\left(B_{{\rm nuc},d}^{z}+\frac{|\vec{B}_{{\rm nuc},d}^{\perp}|^{2}}{2|B_{{\rm ext}}|}\right)S_{d}^{z}.\label{eq: H_el_approx}\end{eqnarray}
The term proportional to $B_{{\rm nuc},d}^z$ is the zeroth-order contribution in $\frac{\left|\vec{B}_{{\rm nuc},d}\right|}{|B_{{\rm ext}}|}$, and may give rise to a magnetic-field-independent contribution to the electron spin decoherence.
The term proportional to $\frac{|\vec{B}_{{\rm nuc},d}^{\perp}|^{2}}{|B_{{\rm ext}}|}$ is the first order correction in $\frac{\left|\vec{B}_{{\rm nuc},d}\right|}{|B_{{\rm ext}}|}$, and is responsible for the interesting collapse-and-revival behavior that we study below.

It should be noted that in writing Eqs.~(\ref{eq:H_z}) and (\ref{eq: H_el_approx}),
we have ignored effects arising from the relative angle $\theta\sim\frac{\left|\vec{B}_{{\rm nuc},d}\right|}{B_{{\rm ext}}}$
between the quantization axes in the two dots, which enter at order
$\left[\frac{\left|\vec{B}_{{\rm nuc},d}\right|}{B_{{\rm ext}}}\right]^{2}$.
First, the misalignment of axes may cause unwanted transitions to
the $\left|T_{\pm}\right\rangle $ states when initializing from the
singlet state, or during free evolution under Hamiltonian $H_{el}$.
However, these effects lead to a reduction of $P_{S}$ on the order
of $\theta^{2}$. Second, we neglect any possible geometric phases
which may accompany the dynamical phase accumulated while the electron
spins adiabatically follow the local quantization axes in their separate
dots. Such phases are proportional to the areas of closed loops swept
out by $\hat{n}_{L}$ and $\hat{n}_{R}$ during their evolution, and
for small $\theta$ are also proportional to $\theta^{2}$.

Although we obtained Eq.~(\ref{eq: H_el_approx}) by treating the
nuclear spin operators as classical vectors in a Taylor series approximation
to Eq.~(\ref{eq:H_z}), an identical expression for $H_{el,z}$ was
formally derived in Ref.~\onlinecite{Cywinski2009} from the full
quantum Hamiltonian, Eq.~(\ref{eq:Hel}), using a Schrieffer-Wolff
transformation. The classical argument above thus provides a simple
intuitive explanation for the formal perturbative derivation. Hereafter,
unless otherwise specified,\textbf{ }we treat the fields in Eq.~(\ref{eq: H_el_approx})
as quantum operators with appropriate commutation relations.

To complete the description of the problem, we now discuss the $\pi$-pulses
employed in the Hahn echo sequence. These pulses are achieved by applying
a time-dependent perturbation $H_{\pi}(t)$, which adds to the system's
total Hamiltonian, $H=H_{el,z}+H_{{\rm nuc}}+H_{\pi}(t)$. We assume
$H_{\pi}(t)$ is only nonzero over narrow intervals which are short
compared to all timescales relevant for evolution under $H_{el,z}$
and $H_{{\rm nuc}}$. Rather than specifying a detailed time-dependent
protocol for $H_{\pi}(t)$, we define $H_{\pi}$ implicitly in terms
of its effect on the electron spin operators $S_{d}^{z}$: \begin{equation}
\tilde{\mathcal{T}}\left[e^{i\int_{0}^{t}H_{\pi}(t')dt'}\right]S_{d}^{z}\,\mathcal{T}\left[e^{-i\int_{0}^{t}H_{\pi}(t'')dt''}\right]=c(t)S_{d}^{z},\label{eq:PiPulse}\end{equation}
 where $\mathcal{T}$ ($\tilde{\mathcal{T}}$) is the (reversed) time-ordering operator. In writing Eq.~(\ref{eq:PiPulse}),
we assume that $H_{\pi}(t)$ acts only within the two-dimensional
qubit subspace. We shall consider perfect $\pi$-pulses, for which
the {}``echo function'' $c(t)$ switches between $1$ and $-1$
over the short duration of the pulse. For simplicity we consider the
pulses to be instantaneous, and for the Hahn echo sequence write \begin{equation}
c(t)=\Theta\left(\tau/2-t\right)-\Theta\left(t-\tau/2\right).\label{eq:EchoFn}\end{equation}

Using Eqs.~(\ref{eq:PiPulse}) and (\ref{eq:EchoFn}) we switch to
an interaction picture with respect to $H_{\pi}(t)$, where 
\begin{equation}
S_{d}^{z}(t)\equiv c(t)S_{d}^{z},
\end{equation}
and where we employ the notation $S_{d}^{z}\equiv S_{d}^{z}\left(t=0\right)$.
The interaction-picture time-dependent Hamiltonian $H_{el,z}(t)$
becomes \begin{equation}
H_{el,z}(t)\approx g^{*}\mu_{B}\sum_{d=L,R}\left(B_{{\rm nuc},d}^{z}+\frac{|\vec{B}_{{\rm nuc},d}^{\perp}|^{2}}{2|B_{{\rm ext}}|}\right)c(t)S_{d}^{z}.\label{eq:H_el_t}\end{equation}
 Equation (\ref{eq:H_el_t}), with Eq.~(\ref{eq:Hnuc}), will serve
as the starting point for our analysis of decoherence in the spin-echo
experiment.

\section{Semi-classical model for the revivals}

\label{sec:simplest_model}

We now show that a simple semi-classical approach in which we treat
the Overhauser field operators as classical vectors can reproduce
the electron spin coherence collapse and revival effect predicted
in Ref.~\onlinecite{Cywinski2009} and observed in Ref.~\onlinecite{ourselves}.
In the next section, we will provide a systematic derivation and justification
of this approach, starting from the full quantum description.

In this section, we treat the Overhauser field in each dot
$d$ in Eq.~(\ref{eq:H_el_t}) as a sum of three classical vectors,
$\vec{B}_{{\rm nuc},d}(t)=\sum_{\alpha=1}^{3}\mathbf{B}_{\alpha,d}(t)$,
where $\alpha$ indexes the three nuclear spin species. We assume
that the magnitudes of the species-dependent fields $\{\left|\mathbf{B}_{\alpha,d}\right|\}$,
and their $z$ components $\{B_{\alpha,d}^{z}\}$, are random but
constant throughout the evolution. 
The time-dependence of $\vec{B}_{{\rm nuc},d}(t)$ within each run arises solely from the Larmor precession of the transverse nuclear spin components.
Explicitly, we neglect the nuclear dipole-dipole interaction, and the influence of the Knight shift on the nuclear Larmor precession.
We assume that 
all nuclei of the same species precess at a single Larmor angular velocity, $\omega_{\alpha}=\gamma_{\alpha}B_{{\rm ext}}$. 

The echo signal $P_{S}(\tau)$ is obtained by averaging over many
experimental runs. Thus we must average the results of electron spin
dynamics against the distribution of initial states of the nuclear
spins. Due to the large number of nuclear spins, $N\approx10^{6}$,
the initial values of the components of each vector $\mathbf{B}_{\alpha,d}(t=0)$
are Gaussian distributed with zero mean and a standard deviation $\overline{b}_{\alpha,d}$
of order $1$ mT (see calculations below). 

The model in Eq.~(\ref{eq:H_el_t}), under the assumptions above,
is sufficient to produce the collapse and revival effect in $P_{S}(\tau)$,
and further provides an intuitive semi-classical picture in which
to understand the phenomenon. However, because we neglect the time
dependence of $B_{{\rm nuc},d}^{z}(t)$, and the effects
of the Knight shift and other dephasing mechanisms of the nuclear
Larmor precession, 
this model does not capture the decaying envelope observed in the
experiment of Ref.~\onlinecite{ourselves}. These issues will be
addressed in detail in Sec.\ref{sec:SC_derivation}.

We now calculate $P_{S}(\tau)$, using the singlet initial state $\left|\psi(t=0)\right\rangle =\left|S\right\rangle =\frac{1}{\sqrt{2}}\left(|\uparrow\downarrow\rangle-|\downarrow\uparrow\rangle\right)$.
For any given set of initial values of the (18 total) components of
the six classical vectors $\left\{ \mathbf{B}_{\alpha,d}(t=0)\right\} $,
the Hamiltonian in Eq.~(\ref{eq:H_el_t}) generates a pure quantum
evolution which after an evolution time $t=\tau$ yields \begin{equation}
\Ket{\psi(\tau)}=\frac{e^{-i\Delta\Phi(\tau)/2}}{\sqrt{2}}\left(\Ket{\uparrow\downarrow}-e^{i\Delta\Phi(\tau)}\Ket{\downarrow\uparrow}\right).\label{eq:psi_f}\end{equation}
 The relative phase $\Delta\Phi(\tau)=\Phi_{L}-\Phi_{R}$ is related
to the difference of the magnitudes of the total effective fields,
$|\vec{B}_{{\rm tot},d}(t)|$, in the two dots. Within the approximation
of $|\vec{B}_{{\rm tot},d}(t)|$ used to write Eq.(\ref{eq:H_el_t}),
we obtain: \begin{equation}
\Phi_{d}(\tau)=\frac{g^{*}\mu_{B}}{2|B_{{\rm ext}}|}\int_{0}^{\tau}|\vec{B}_{{\rm nuc},d}^{\perp}(t)|^{2}c(t)\, dt.\label{eq: phi_LR}\end{equation}
 Note that the phase $\Phi_{d}$ is determined solely by the dynamics
of a single isolated electron in dot $d$. Below we will use this
fact to relate the decoherence of the two-electron singlet-triplet
qubit to that of a single electron spin in a quantum dot.

For the final state $\Ket{\psi(\tau)}$ in Eq.~(\ref{eq:psi_f}),
the singlet return probability is given by \begin{equation}
|\Amp{S}{\psi(\tau)}|^{2}=\frac{1}{2}+\frac{1}{2}\cos(\Delta\Phi).\label{eq: P_no_everage}\end{equation}
 The ensemble-averaged singlet return probability $P_{S}(\tau)$ is
found by averaging the result for a single run, Eq.~(\ref{eq: P_no_everage}),
with respect to the distribution of initial magnitudes and directions
of the six vectors $\{\mathbf{B}_{\alpha,d}(t=0)\}$. Note that because
$\cos(\Delta\Phi)={\rm Re}\left[e^{i\Phi_{R}}e^{-i\Phi_{L}}\right]$,
and because the Overhauser field configurations in the two dots are
independent, we can average over $e^{i\Phi_{L}}$ and $e^{i\Phi_{R}}$
independently, \begin{equation}
P_{s}=\frac{1}{2}+\frac{1}{2}{\rm Re}\Big[\langle e^{i\Phi_{R}}\rangle\langle e^{-i\Phi_{L}}\rangle\Big].\label{eq:Ps_0}\end{equation}
 To perform the averaging, we calculate $\Phi_{d}(\tau)$ using Eq.~(\ref{eq: phi_LR}),
with $c(t)$ given by Eq.~(\ref{eq:EchoFn}), and with the classical
evolution for $\vec{B}_{{\rm nuc},d}^{\perp}(t)$ resulting from the
free precession of the underlying nuclear spins. We write $|\vec{B}_{{\rm nuc},d}^{\perp}(t)|^{2}=\sum_{\alpha,\beta}b_{\alpha,d}(t)b_{\beta,d}^{*}(t)$
in terms of the complex-valued fields $b_{\alpha,d}=B_{\alpha,d}^{x}+iB_{\alpha,d}^{y}$.
In this decomposition, the time evolution of the Overhauser field
due to Larmor precession is given by $b_{\alpha,d}(t)=b_{\alpha,d}(0)e^{i\omega_{\alpha}t}$.
Each phase $\Phi_{d}$ is then given by \begin{eqnarray}
\Phi_{d} & = & \frac{g^{*}\mu_{B}}{2|B_{{\rm ext}}|}\int_{0}^{\tau}c(t)\sum_{\alpha,\beta}b_{\alpha,d}(t)b_{\beta,d}^{*}(t)\, dt\label{eq:Phi1}\\
 & = & \frac{2g^{*}\mu_{B}}{|B_{{\rm ext}}|}\sum_{\alpha,\beta}b_{\alpha,d}(0)b_{\beta,d}^{*}(0)\frac{e^{i\omega_{\alpha\beta}\tau/2}}{i\omega_{\alpha\beta}}\sin^{2}\left(\omega_{\alpha\beta}\tau/4\right),\nonumber \end{eqnarray}
 where $\omega_{\alpha\beta}=\omega_{\alpha}-\omega_{\beta}$.

Next we must integrate over all initial conditions, i.e. over the
initial magnitudes and phases of the three fields $\{b_{\alpha,d}(0)\}$
in each dot $d$. For this purpose we express the initial conditions
as $b_{\alpha,d}(0)=\overline{b}_{\alpha,d}z_{\alpha}$, where each
$z_{\alpha}=x_{\alpha}+iy_{\alpha}$ is a dimensionless complex variable.
The quantity $\overline{b}_{\alpha,d}$ is the root-mean-squared (rms)
value of each component of the transverse field for species $\alpha$
in dot $d$, 

\begin{equation}
g^*\mu_B\overline{b}_{\alpha,d}=\sqrt{a_{\alpha}\bar{n}_{\alpha}/N_{d}}\mathcal{A}_{\alpha},\label{eq:b_alpha}
\end{equation}

with $a_{\alpha}=\frac{2}{3}(I_{\alpha}+1)I_{\alpha}=\frac{5}{2}$.
Here $\bar{n}_{\alpha}$ is the average number of nuclei of species
$\alpha$, per unit cell (see Ref.~\onlinecite{Cywinski2009}).
This substitution gives \begin{eqnarray}
\Phi_{d} & = & \sum_{\alpha,\beta}T_{\alpha\beta,d}\frac{z_{\alpha}z_{\beta}^{*}}{2},\label{eq: T_alpha_beta0}\end{eqnarray}
 with \begin{equation}
T_{\alpha\beta,d}=-\frac{4g^{*}\mu_{B}\overline{b}_{\alpha,d}\overline{b}_{\beta,d}}{|B_{{\rm ext}}|}\frac{e^{i\omega_{\alpha\beta}\tau/2}}{i\omega_{\alpha\beta}}\sin^{2}\left(\omega_{\alpha\beta}\tau/4\right).\label{eq:T_alpha_beta}\end{equation}

We now carry out an ensemble average over the initial conditions by
treating all components of the $\{z_{\alpha}\}$ as independent Gaussian-distributed
random variables with zero mean and unit variance, i.e.~with probability
density function $p(\{z_{\alpha'}\})\prod_{\alpha=1}^{3}dx_{\alpha}dy_{\alpha}=\frac{1}{\left(2\pi\right)^{3}}\prod_{\alpha=1}^{3}\exp(-\frac{|z_{\alpha}|^{2}}{2})dx_{\alpha}dy_{\alpha}$:
\begin{eqnarray}
\langle e^{-i\Phi_{d}}\rangle & = & \int\prod_{\alpha}{dx_{\alpha}dy_{\alpha}\, p(\{z_{\alpha'}\})\, e^{-\frac{i}{2}\sum_{\beta,\beta'}T_{\beta\beta',d}z_{\beta}^{*}z_{\beta'}}}\nonumber \\
 & = & \prod_{\alpha}\frac{1}{1+i\lambda_{\alpha,d}}.\label{eq:eiphi1}\end{eqnarray}
 Here the parameters $\{\lambda_{\alpha,d}\}$ are the eigenvalues
of the $T$-matrix for dot $d$.

The $3\times3$ Hermitian matrix $T$ in Eq.~(\ref{eq:T_alpha_beta})
corresponds to that of Ref.~\onlinecite{Cywinski2009}. 
Because $T$ is Hermitian and is similar to an antisymmetric matrix, it has one zero eigenvalue, $\lambda_{1,d}=0$, and a pair of eigenvalues $\lambda_{2(3),d}=\pm\sqrt{\sum_{\alpha>\beta}{\left|T_{\alpha\beta,d}\right|^{2}}}$.
Inserting these eigenvalues into Eq.~(\ref{eq:eiphi1}), and using
Eq.~(\ref{eq:T_alpha_beta}) for $T_{\alpha\beta,d}$, we obtain:
\begin{equation}
\langle e^{-i\Phi_{d}}\rangle=\left[1+\sum_{\alpha>\beta}\left(\frac{4g^{*}\mu_{B}\overline{b}_{\alpha,d}\overline{b}_{\beta,d}}{|B_{{\rm ext}}|}\right)^{2}\frac{\sin^{4}\left(\omega_{\alpha\beta}\tau/4\right)}{\omega_{\alpha\beta}^{2}}\right]^{-1}.\label{eq:eiphi2}\end{equation}
 Thus we see that 
the semi-classical model used in this section reproduces the result of 
Ref.~\onlinecite{Cywinski2009}
for the decay of the spin echo signal in a single quantum dot. 
The echo signal shows oscillations with amplitude $\left(\frac{4g^{*}\mu_{B}\overline{b}_{\alpha,d}\overline{b}_{\beta,d}}{|B_{{\rm ext}}|\omega_{\alpha\beta}}\right)^{2}$, which develop into the complete collapses and revivals at low magnetic fields. Note that the expression for the phase in Eq.~(\ref{eq:Phi1}) includes terms which are bilinear combinations of Gaussian variables, and so are not Gaussian themselves. 
Therefore the decay in the interval $\tau\ll1/\omega_{\alpha\beta} $ behaves like an inverse polynomial, rather than the form $e^{-{\rm const}\cdot\tau^4}$ expected for spectral diffusion (see below and Refs.~\onlinecite{Witzel2006,Witzel2008,Sousa2009}).


We now return to computing the echo signal in the double dot system, Eq.~(\ref{eq:Ps_0}).
First, note that $\langle e^{-i\Phi_{d}}\rangle$ in Eq.~(\ref{eq:eiphi2})
is strictly real. This fact is a consequence of antisymmetry of the echo function 
around the time $\tau/2$. Thus we can drop the ``${\rm Re}$''
from Eq.~(\ref{eq:Ps_0}) and write the echo signal as\begin{equation}
2P_{S}-1=\langle e^{-i\Phi_{L}}\rangle\langle e^{-i\Phi_{R}}\rangle.\label{eq:echo1}\end{equation}

The expressions in Eqs.~(\ref{eq:eiphi1}) and (\ref{eq:eiphi2}),
which describe the dephasing of a single electron spin in an isolated
quantum dot, were derived previously from a fully quantum mechanical
treatment in Ref.~\onlinecite{Cywinski2009}. A key element of
the derivation in that work was the vanishing of the contribution
of the commutator $[I_{k}^{+},I_{l}^{-}]$ between nuclear spin operators
in the low order perturbation expansion of the evolution operator.
The vanishing commutator is indicative of classical behavior, and
further motivates our classical treatment of the nuclear evolution.

To conclude this section, we show that the classical treatment provides
an intuitive explanation for the collapses and revivals. The total
effective electron Zeeman field in each dot d, $\left|\mathbf{B}_{{\rm ext}}+\mathbf{B}_{{\rm nuc},d}\right|$,
depends on the square of the transverse Overhauser field, $|\vec{B}_{{\rm nuc},d}^{\perp}|^{2}$.
The terms in the expansion $|\vec{B}_{{\rm nuc},d}^{\perp}|^{2}=\sum_{\alpha\alpha'}\left(B_{\alpha,d}^{x}B_{\alpha',d}^{x}+B_{\alpha,d}^{y}B_{\alpha',d}^{y}\right)$
which involve nuclear spins of two different species, $\alpha\neq\alpha'$,
oscillate at the relative Larmor angular velocity $\omega_{\alpha}-\omega_{\alpha'}$.
As a result, the magnitude of the total field, $|\vec{B}_{{\rm tot},d}|$,
and hence the splitting between electron spin energy levels, oscillates
as a function of time. These oscillations determine the phase accumulation
during each run of the experiment. In this simple semi-classical treatment the nuclear spin evolution is not affected by the electron spin, so that the phase between electron spin components remains well-defined during each run. However, averaging over the ensemble of nuclear spin initial states amounts to averaging over the phases and amplitudes
of these oscillations, and causes the collapse of the echo signal.
However, if the free evolution time $\tau/2$ is simultaneously an
integer multiple of each of the three relative Larmor periods, then
the contribution of those oscillations vanishes independently of the
initial nuclear state, and a revival of electron spin coherence is
observed. Note that it is a fortunate coincidence in GaAs that the
Larmor frequencies of the three species are nearly equidistant, which,
to a good approximation, allows the commensurability condition to
be easily fulfilled simultaneously for all three pairs of nuclear
species.

\section{Derivation of the semi-classical approach }

\label{sec:SC_derivation}

In this section we present a systematic derivation of the semi-classical
treatment of electron spin dynamics presented above. Keeping in mind
the discussion surrounding Eq.~(\ref{eq: H_el_approx}), we now restore
the quantum nature of the electron spin operators and begin with the
interaction-picture quantum Hamiltonian $H=H_{el,z}(t)+H_{{\rm nuc}}$,
Eq.~(\ref{eq:H_el_t}). We also now reintroduce the nuclear dipolar interaction, see Eq.~(\ref{eq:Hnuc}), and allow for local variations of the nuclear Larmor frequencies, 
which were omitted from the simple model in section \ref{sec:simplest_model}.
At each step in the derivation, we will justify the approximations
needed to arrive at the semi-classical model with arguments from first
principles, or with the help of experimental data. Along the way,
we will also incorporate several important features of the dynamics,
such as the time dependence of $B_{{\rm nuc}}^{z}(t)$, back-action
of the electron spin evolution on the nuclear state, and nuclear spin
dephasing, which were ignored in the heuristic treatment in the previous
section.

\subsection{Quantum expression for the echo signal}

Our aim is to derive an expression for the echo signal at the end
of the full Hahn echo sequence, via the calculation of the singlet
return probability \begin{equation}
P_{S}(\tau)=\frac{1}{Z_{\infty}}{\rm Tr}_{{\rm nuc}}\Big[\left\langle S\right|U^{\dagger}(\tau)\left|S\right\rangle \left\langle S\right|U(\tau)\left|S\right\rangle \Big],\label{eq:P_S}\end{equation}
 where $U(\tau)=\mathcal{T}e^{-i\int_{0}^{\tau}{\left(H_{el,z}(t)+H_{{\rm nuc}}\right)dt}}$
is the interaction picture evolution operator of the joint electron-nuclear
spin system, with $\mathcal{T}$ representing the time-ordering operator.
Here we have assumed that the electron spins are initialized to the
singlet state $\Ket{S}$. The trace in Eq.~(\ref{eq:P_S}) is taken
over all nuclear spin states, with $Z_{\infty}={\rm Tr}_{{\rm nuc}}[\hat{1}]$
representing the partition function for an infinite temperature (completely
random) nuclear spin state.

To separate the time evolution due to the static part of $B_{{\rm nuc}}^{z}$
(which dominates the evolution for $t<\tau$, but has no influence
on $U(t=\tau)$ due to the spin-echo), we introduce the zero-order
Hamiltonian \begin{equation}
H_{0}(t)=g^{*}\mu_{B}\sum_{d=L,R}B_{{\rm nuc},d}^{z}\, c(t)S_{d}^{z}+\sum_{k}\omega_{k}I_{k}^{z},\label{H0}\end{equation}
 and a corresponding zero-order evolution operator $U_{0}(t)=e^{-i\int_{0}^{t}H_{0}(t')dt'}$.
Note that due to antisymmetry of the echo function, $\int_{0}^{\tau}c(t')dt'=0$,
the evolution operator $U_{0}(t=\tau)$ at the end of the full sequence
does not depend on the electron spin operators and simply rotates
all nuclear spins about the $z$-axis.

The full evolution operator $U(\tau)$ can be rewritten as \begin{equation}
U(\tau)=U_{0}(\tau)\cdot\mathcal{T}e^{-i\int_{0}^{\tau}{dt\sum_{d}\left[H_{\perp,d}(t)S_{d}^{z}c(t)+H_{D,d}(t)\right]}},\label{eq:U1}\end{equation}
 where the time-dependent operators in the exponent are given by $H_{\perp,d}(t)=U_{0}^{\dagger}(t)\left[g^{*}\mu_{B}\frac{|\vec{B}_{{\rm nuc},d}^{\perp}|^{2}}{2|B_{{\rm ext}}|}\right]U_{0}(t)$
and $H_{D,d}(t)=U_{0}^{\dagger}(t)\Big[\sum_{n,n'\in d}{D_{n,n'}^{ij}I_{n}^{i}I_{n'}^{j}}\Big]U_{0}(t)$,
see Eq.~(\ref{eq:Hnuc}). These terms describe the evolution of the
Overhauser fields due to Larmor precession and the dipole-dipole interaction
between nuclear spins in dot $d$, respectively. We assume that the
two dots are well separated, such that the inter-dot dipolar coupling
can be neglected. This approximation is not essential for the derivation
ahead, however, and the existence of a small inter-dot coupling would
not significantly affect the final result.

In order to evaluate Eq.~(\ref{eq:P_S}), we decompose the evolution
operator $U(\tau)$ into four separate pieces. Because operators describing
spins in different dots commute, the exponentials in Eq.~(\ref{eq:U1})
can be factored by dot index $d$. Therefore the evolution operator
can be written as $U(\tau)=U_{L}(\tau)U_{R}(\tau)$, where $U_{L(R)}(\tau)$
only involves spin operators in the left (right) dot. In addition,
because the $z$-projection of electron spin in each dot is conserved
by $U(\tau)$, see Eq.~(\ref{eq:U1}), we can separate the evolution
by introducing projectors onto the product states $\Ket{\uparrow\downarrow}$
and $\Ket{\downarrow\uparrow}$: \begin{equation}
U(\tau)=U_{L+}(\tau)U_{R-}(\tau)\left|\uparrow\downarrow\right\rangle \left\langle \uparrow\downarrow\right|+U_{L-}(\tau)U_{R+}(\tau)\left|\downarrow\uparrow\right\rangle \left\langle \downarrow\uparrow\right|,\label{eq:U_spin}\end{equation}
 where $U_{d\sigma}$ is a unitary operator acting only on the nuclear
spins in dot $d$, with the electron spin taken to be in the $S_{d}^{z}$
eigenstate $\Ket{\uparrow}$ for $\sigma=+$, or $\Ket{\downarrow}$
for $\sigma=-$. 
Inserting Eq.~(\ref{eq:U_spin}) into Eq.~(\ref{eq:P_S}), and performing
the trace over nuclear spin states separately for the two dots, we
find that the echo signal is given by {[}\textit{c.f.} Eq.~(\ref{eq:echo1}){]}
\begin{equation}
2P_{S}-1={\rm Re}\left(C_{L}\cdot C_{R}\right),\label{eq:PS2}\end{equation}
 where \begin{equation}
C_{d}=\frac{1}{Z_{\infty,d}}{\rm Tr}_{{\rm nuc}}\left[U_{d+}^{\dagger}(\tau)U_{d-}(\tau)\right],\: d=L,R\label{eq:C_d}\end{equation}

The echo signal, Eq.~(\ref{eq:PS2}), is a product of two similar
averages which are taken over the different sets of nuclear spins
in the two dots. In the remainder of this section we focus on the
behavior of the average within a single dot, which we denote as $C=\frac{1}{Z_{\infty}}{\rm Tr}_{{\rm nuc}}\left[U_{+}^{\dagger}(\tau)U_{-}(\tau)\right]$.
This allows us to simplify all formulas by suppressing the dot index
$d$. At the end of subsection \ref{sub: calculation} we will return
to the two-electron double-dot echo signal, including the combined
effects of dephasing in each of the two dots.

\subsection{The semi-classical approximation of separating the dynamics of the
spin diffusion and transverse Overhauser field}

In this subsection we introduce a semi-classical treatment in which
the evolution operators in Eq.(\ref{eq:PS2}) are factorized into
contributions depending separately on $H_{\perp}(t)$ and $H_{D}(t)$.
We start by using Eq.~(\ref{eq:U1}) and the fact that $U_{0}(\tau)$
involves only nuclear spin operators to write the evolution operator
in $C$ as \begin{eqnarray}
U_{+}^{\dagger}(\tau)U_{-}(\tau) & = & \widetilde{\mathcal{T}}\!\left[e^{i\int_{0}^{\tau}{dt\,\left[\frac{1}{2}H_{\perp}^{+}(t)c(t)+H_{D}^{+}(t)\right]}}\right]\label{eq:Ud_upUd_dwn}\\
 & \times & \!\mathcal{T}\left[e^{-i\int_{0}^{\tau}{dt'\left[-\frac{1}{2}H_{\perp}^{-}(t')c(t')+H_{{D}}^{-}(t')\right]}}\right],\nonumber \end{eqnarray}
where the superscripts $+$ and $-$ indicate projection
onto the $\Ket{\uparrow}$ and $\Ket{\downarrow}$ electron spin states.
The projection is needed because $H_{\perp}(t)$ and $H_{D}(t)$ depend
on $U_{0}(t)$, which for $t\neq\tau$ depends on $S^{z}$.

Intuitively, the two terms involving $H_{\perp}(t)$ and $H_{D}(t)$
lead to suppression of the echo signal, each through a different physical
mechanism. In the spirit of the semi-classical approximation in Sec.~\ref{sec:simplest_model},
where the Overhauser field operators were treated as classical vector
variables, $H_{\perp}(t)$ causes dephasing through the time-dependence
of $|\vec{B}_{{\rm nuc}}^{\perp}(t)|$ generated by nuclear Larmor
precession, while $H_{D}(t)$ causes decoherence through fluctuations
of $B_{{\rm nuc}}^{z}(t)$ generated by dipolar-interaction-mediated
spin diffusion.

The semi-classical approach relies on three main approximations, which
involve neglecting various commutators of the form $[B_{{\rm nuc}}^{i}(t),B_{{\rm nuc}}^{j}(t')]$.
When the number of nuclei $N_{d}$ is large, these commutators scale
as $|\vec{B}_{{\rm nuc}}(t)|/N_{d}$. Below we provide a physical
motivation for each such approximation, and discuss the associated
range of validity. 

First, we neglect the commutator between $H_{\perp}(t)$ and $H_{D}(t')$,
and factor the exponentials appearing in Eq.~(\ref{eq:Ud_upUd_dwn}):
\begin{equation}
U_{\sigma}(\tau)\approx U_{\perp\sigma}(\tau)U_{D\sigma}(\tau),\label{eq:sep_approx}\end{equation}
 with \begin{eqnarray}
U_{\perp\sigma}(\tau) & = & \mathcal{T}e^{-\frac{i}{2}\int_{0}^{\tau}{\sigma H_{\perp}^{\sigma}(t)c(t)}dt},\label{Udperp}\\
U_{D\sigma}(\tau) & = & \mathcal{T}e^{-i\int_{0}^{\tau}{H_{D}^{\sigma}(t)}dt}.\nonumber \end{eqnarray}
 The commutator $[H_{\perp}(t),H_{D}(t')]$ which was set to zero
in writing Eq.~(\ref{eq:sep_approx}) leads to nuclear spin dephasing,
i.e. decay of the correlator $\langle I_{k}^{i}(t)I_{k}^{j}(t')\rangle$,
due to dipolar flip-flop events. In our model, we account for this
effect phenomenologically by introducing random site-to-site small
variations of the nuclear Larmor frequencies (see Sec.~\ref{Sec:Avg}).
In doing so, we assume that intrinsic and extrinsic dephasing of nuclear
spins affect the electron spin dynamics in the same way.

In addition, the commutators neglected above may lead to
enhanced nuclear spin diffusion through the combination of dipole-dipole
and hyperfine-mediated spin flips. The validity of this approximation
thus depends on material and system parameters. However, it can be
checked by comparison with experimental data, as discussed in more
detail in section.~\ref{sec:Discussion}.

The second approximation is to split the average of the product of
operators involving $U_{D\sigma}$ and $U_{\perp\sigma}$ into a product
of averages involving $U_{D\sigma}$ and $U_{\perp\sigma}$ separately:
\begin{widetext} \begin{equation}
\frac{1}{Z_{\infty}}{\rm Tr}_{{\rm nuc}}\left[U_{D+}^{\dagger}U_{\perp+}^{\dagger}U_{D-}U_{\perp-}\right]\approx\frac{1}{Z_{\infty}}{\rm Tr}_{{\rm nuc}}\left[U_{D+}^{\dagger}U_{D-}\right]\cdot\frac{1}{Z_{\infty}}{\rm Tr}_{{\rm nuc}}\left[U_{\perp+}^{\dagger}U_{\perp-}\right].\label{eq:prod_averages}\end{equation}
 \end{widetext} The correlations which are neglected by splitting
this average are related to correlations of the longitudinal and transverse
components of the Overhauser fields, $\vec{B}_{{\rm nuc}}^{z}(t)$
and $\vec{B}_{{\rm nuc}}^{\perp}(t)$. These correlations are contributed
by expectation values of at least four operators of the same nuclear
spin $n$, e.g.~$\left(I_{n}^{x}\right)^{2}\left(I_{n}^{z}\right)^{2}$.
On the other hand, the leading contribution to the split-average comes
from the expectation values of only two operators of the same nuclear
spin, e.g.~$\left(I_{n}^{z}\right)^{2}$, and therefore is roughly
$N_{d}$ times bigger, due to the fact that it contains a sum over
at least $N_{d}$ times as many terms.\textbf{ }

\subsection{The semi-classical approximation of averaging over all nuclear spin
states\label{sub: calculation}}

\label{Sec:Avg}

Through Eq. (\ref{eq:prod_averages}), the dephasing in a single dot
can be approximately expressed as a product of two separate averages,
one involving $U_{D\sigma}$ and the other involving $U_{\perp\sigma}$.
The average over $U_{D\sigma}$ can be associated with decoherence
due to spectral diffusion caused by dipole-dipole mediated nuclear
spin diffusion. This process was analyzed in Refs.~\onlinecite{Witzel2006,Witzel2008,Sousa2009},
and was shown to result in an echo decay factor $\exp(-(\tau/\tau_{SD})^{4})$.
Using this result and Eq.~(\ref{eq:prod_averages}), we rewrite Eq.~(\ref{eq:C_d})
as \begin{equation}
C\left(\tau\right)\approx e^{-(\tau/\tau_{SD})^{4}}\frac{1}{Z_{\infty}}{\rm Tr}_{{\rm nuc}}\left[U_{\perp+}^{\dagger}(\tau)U_{\perp-}(\tau)\right],\label{eq:PS3}\end{equation}
 The “spectral diffusion time” $\tau_{SD}$ depends on the details of the quantum dots, and weakly depends on the magnetic field strength. According to Refs. \onlinecite{Witzel2006,Witzel2008} $\tau_{SD}$ is of order 10 microseconds, which is comparable to the duration of the experiment. Thus, the decay factor is likely to be important for a detailed fit to the experiments, and will be dominant at high magnetic fields. However, as we show below, at low enough magnetic field the dephasing associated with $U_{\perp\sigma}$, i.e.~with the evolution of the transverse components of the Overhauser
fields, becomes dominant.

We now turn to calculate the the dephasing associated with $U_{\perp\sigma}$ semi-classically. Making use of the large number of nuclei involved, $N_{d}\gtrsim10^{6}$,
the trace in Eq.~(\ref{eq:PS3}) can be cast into a form which closely
resembles the classical average described in Sec.~\ref{sec:simplest_model}.
We first combine together large groups of nuclei with similar couplings
to form a set of {}``giant'' collective spins. Then, by evaluating
the trace in a basis of coherent states with well defined orientations
of these giant spins, we find that their induced Overhauser fields
effectively act as classical variables like those introduced {}``by
hand'' in Sec.~\ref{sec:simplest_model}. However, there are two
important differences in this more refined treatment. First, whereas
in Sec.~\ref{sec:simplest_model} all nuclear spins within each species
were forced to precess at a fixed Larmor frequency, we now account
for the electron-spin-dependent shift (i.e.~the Knight shift) of
the nuclear precession rate which is inherent in Eqs.~(\ref{eq: H_el_approx})
and (\ref{H0}). Second, we now account for inhomogeneity in the system,
both in the Knight field terms and in the Larmor frequencies of all
nuclear spins. Site-to-site variations of the Larmor frequencies $\omega_{n}$,
see Eq.~(\ref{eq:Hnuc}), phenomenologically account for nuclear
spin dephasing due to, e.g. local quadrupole moments. We assume that
the inhomogeneities are weak, such that the differences between the
mean Larmor angular velocities of the three species, $\omega_{\alpha}=\gamma_{\alpha}B_{{\rm ext}}$,
are much larger than the widths of the distributions for each one.

To account for inhomogeneities in the system, we divide the nuclei
in each dot into $K$ groups labeled by the index $k=1,\ldots,K$,
with each group $k$ containing nuclei of the same species $\alpha(k)$.
These $\tilde{N}_{k}$ are picked from the $N_{d}$ nuclei in dot
$d$ such that all members of the group have nearly the same hyperfine
coupling $A_{k}$ and feel nearly the same phenomenological local
shift in magnetic field, $\delta B_{k}$. The Larmor angular velocity
for nuclear spins in group $k$ is given by $\omega_{\alpha(k)}+\delta\omega_{k}+\frac{1}{2}\sigma c(t)A_{k}$
where $\delta\omega_{k}=\gamma_{\alpha(k)}\delta B_{k}$. Note that
the sign of the Knight shift $\pm\frac{1}{2}A_{k}$ depends on the
state of the electron spin in the dot at time $t$, and is therefore
proportional to $\sigma c(t)$.

To calculate $C$, the single dot contribution to the echo signal,
we need to choose a basis for the nuclear spin Hilbert space. For
each group $k$, we form a collective spin from all of the
nuclear spins in the group, and consider ``giant spin'' states
of well-defined total angular momentum $I_{k}$ and orientation $\hat{\vec{n}}_{k}=(\sin\theta_{k}\cos\phi_{k},\,\sin\theta_{k}\sin\phi_{k},\,\cos\theta_{k})$:
$\left|\mathbf{I}_{k}\right|^{2}\Ket{I_{k},\hat{\vec{n}}_{k}}=I_{k}(I_{k}+1)\Ket{I_{k},\hat{\vec{n}}_{k}}$,
$(\hat{\vec{n}}_{k}\cdot\vec{I}_{k})\Ket{I_{k},\hat{\vec{n}}_{k}}=I_{k}\Ket{I_{k},\hat{\vec{n}}_{k}}$,
where $\theta_{k}$ and $\phi_{k}$ are the polar and azimuthal angles
of giant spin $k$, respectively. A wave function in the Hilbert space
of all nuclear spins is written as $\left|\Psi\right\rangle =\bigotimes_{k=1}^{K}\left|I_{k},\hat{\vec{n}}_{k}\right\rangle $.
The trace is then performed by summing over all possible assignments
of the values $\{I_{k}\}$ with appropriate weights\cite{Mikhailov1977},
and integrating over all possible directions $\{\hat{\vec{n}}_{k}\}$.
The length $I_{k}$ of each giant spin $k$ can vary from $0$ to
$\frac{3}{2}\tilde{N}_{k}$, but the vast majority of giant spin states have
lengths of order $I_{k}\sim\sqrt{\tilde{N}_{k}}$. The relative quantum
uncertainty in the transverse components of the giant spin coherent
states, $\Delta I_{k}^{i}/I_{k}^{i}$, scales as $1/\sqrt{I_{k}}=1/\sqrt[4]{\tilde{N}_{k}}$.
As we shell see below, for analyzing the behavior near the revivals
peaks it is sufficient to divide the nuclei into about $K\sim10$
groups, which leaves as many as $\tilde{N}_{k}\approx10^{5}$ nuclei in
each group.

In the coherent state basis, the semi-classical approximation converts
the expectation values $\left\langle \Psi\right|U_{\perp d+}^{\dagger}(\tau)U_{\perp d-}(\tau)\left|\Psi\right\rangle $,
which arise in the evaluation of the trace in Eq.~(\ref{eq:PS3}),
into {}``classical'' expressions \begin{widetext} \begin{equation}
\left\langle \Psi\right|U_{\perp+}^{\dagger}(\tau)U_{\perp-}(\tau)\left|\Psi\right\rangle \simeq e^{-i\int_{0}^{\tau}\frac{g^{*}\mu_{B}}{4B_{{\rm ext}}}|\vec{B}_{{\rm nuc},d}^{\perp+}(t)|_{cl}^{2}c(t)dt}e^{-i\int_{0}^{\tau}\frac{g^{*}\mu_{B}}{4B_{{\rm ext}}}|\vec{B}_{{\rm nuc},d}^{\perp-}(t)|_{cl}^{2}c(t)dt},\label{eq:semi_classical_approx}\end{equation}
 where $|\vec{B}_{{\rm nuc},d}^{\perp+}(t)|_{cl}^{2}$ is determined
by the expectation values $\overline{I_{k}^{\sigma x(y)}(t)}=\MatEl{I_{k},\hat{\vec{n}}_{k}}{I_{k}^{\sigma x(y)}(t)}{I_{k},\hat{\vec{n}}_{k}}$
of the giant spin components:

\[
\frac{g^{*}\mu_{B}}{4B_{{\rm ext}}}|\vec{B}_{{\rm nuc},d}^{\perp\sigma}(t)|_{cl}^{2}\equiv\frac{1}{4g^{*}\mu_{B}B_{{\rm ext}}}\sum_{k,k'}A_{k}A_{k'}\left(\overline{I{}_{k}^{\sigma x}(t)}\cdot\overline{I_{k'}^{\sigma x}(t)}+\overline{I_{k}^{\sigma y}(t)}\cdot\overline{I_{k'}^{\sigma y}(t)}\right).\]

\end{widetext} The superscript $\sigma$ indicates that the time
dependence of the operators is determined by the evolution with respect
to $H_{0}$, Eq. (\ref{H0}), which depends on the electron spin state
$\sigma$.

The validity of the approximation, Eq.~(\ref{eq:semi_classical_approx}),
is discussed in Sec.~\ref{sub:quantum error} below. After making
the approximation, however, the trace reduces to an integration over
all possible expectation values $\overline{I_{k}^{\sigma x(y)}(t)}$
of the components of the $K$ giant spins. We perform the integrations
using a similar method to that used in Section \ref{sec:simplest_model}.
First we write the classical field $|\vec{B}_{{\rm nuc}}^{\perp\sigma}(t)|_{cl}^{2}=\sum_{k,l}b_{k}^{\sigma}(t)b_{l}^{\sigma*}(t)$
as a sum of products of complex variables, where $b_{k}^{\sigma}=A_{k}\left(\overline{I{}_{k}^{\sigma x}(t)}+i\overline{I{}_{k}^{\sigma y}(t)}\right)$.
Each $b_{k}^{\sigma}(t)$ evolves according to $b_{k}^{\sigma}(t)=b_{k}(0)e^{i[\omega_{\alpha(k)}t+\delta\omega_{k}t+\sigma A_{k}\int_{0}^{t}c(t')dt']}$.
Using the approximation in Eq.~(\ref{eq:semi_classical_approx})
to evaluate the trace in Eq.~(\ref{eq:PS3}), we have \begin{equation}
C\left(\tau\right)\approx C_{sc}\left(\tau\right)=e^{-(\tau/\tau_{SD})^{4}}\left\langle e^{i\Phi}\right\rangle ,\label{eq: quantum2classical}\end{equation}
 where the angled brackets indicate a {}``classical'' averaging
over the complex random variables $b_{k}^{\sigma}(0)$, and the phase
$\Phi$ is given by the integrals in the exponents in Eq.~(\ref{eq:semi_classical_approx})
\begin{widetext} \begin{eqnarray}
\Phi\left(\tau\right) & = & \frac{g^{*}\mu_{B}}{4|B_{{\rm ext}}|}\sum_{\sigma=\pm1}\int_{0}^{\tau}dt\sum_{k,l}b_{k}^{\sigma}(t)b_{l}^{\sigma*}(t)\nonumber \\
 & = & \frac{g^{*}\mu_{B}}{4|B_{{\rm ext}}|}\sum_{k,l}b_{k}(0)b_{l}(0)\sum_{\sigma=\pm1}\int_{0}^{\tau}dt\, c(t)\exp\left[i(\omega_{kl}+\delta\omega_{kl})t+i\sigma A_{kl}\int_{0}^{t}dt'c(t')\right],\label{eq:Tkl0}\end{eqnarray}
 \end{widetext} with $\omega_{kl}=\omega_{\alpha(k)}-\omega_{\alpha(l)}$,
$\delta\omega_{kl}=\delta\omega_{k}-\delta\omega_{l}$, and $A_{kl}=A_{k}-A_{l}$.

Assuming that $\tilde{N}_{k}$ is large for every group $k$, we perform
the average over all initial coherent nuclear spin states by writing
$b_{k}(0)=\overline{b}_{k}z_{k}$ in terms of a collection of independent,
complex Gaussian random variables $\{z_{k}\}$ with unit variance.
Here $g^*\mu_B\overline{b}_{k}=\sqrt{\widetilde{N}_{k}a_{\alpha(k)}/2}A_{k}$
is the rms value of each component of the Overhauser field associated
with group $k$ in the dot, similar to the parameters $\{\overline{b}_{\alpha}\}$
appearing in Eq.~(\ref{eq:T_alpha_beta}). In this representation,
Eq.~(\ref{eq:Tkl0}) becomes \begin{equation}
\Phi\left(\tau\right)=\sum_{k,l}T_{kl}\frac{z_{k}z_{l}^{*}}{2},\label{eq: Phi_d_zz}\end{equation}
 with \begin{eqnarray}
T_{kl}\left(\tau\right) & = & \frac{ig^{*}\mu_{B}\overline{b}_{l}\overline{b}_{k}(\omega_{kl}+\delta\omega_{kl})}{2|B_{{\rm ext}}|}\label{Tkld}\\
 & \times & 4\frac{\cos(A_{kl}\tau/2)-\cos\left[(\omega_{kl}+\delta\omega_{kl})\tau/2\right]}{(\omega_{kl}+\delta\omega_{kl})^{2}-A_{kl}^{2}}.\nonumber \end{eqnarray}

Performing the Gaussian average over $\{z_{k}\}$, we obtain \begin{eqnarray}
\langle e^{-i\Phi}\rangle & = & \int\prod_{k'}dx_{k'}dy_{k'}p(|z_{k'}|^{2})\exp\left(-i\sum_{k,l}T_{kl}\frac{z_{k}^{*}z_{l}}{2}\right)\nonumber \\
 & = & \prod_{m}\frac{1}{1+i\lambda_{m}\left(\tau\right)},\label{eq:eiphi3}\end{eqnarray}
 where the parameters $\{\lambda_{m}\}$ are the eigenvalues of the
$M\times M$ Hermitian matrix with elements $\{T_{kl}\}$.

Finally, we calculate the spin-echo signal of the two-electron double
quantum dot singlet-triplet qubit, incorporating the dephasing due
to both two dots, Eq. (\ref{eq:PS2}), using Eqs.~(\ref{eq: quantum2classical})
and (\ref{eq:eiphi3}) for each dot $d=L,R$: \begin{equation}
2P_{s}-1=e^{-(\tau/\tilde{\tau}_{SD})^{4}}\prod_{d=L,R}\left(\prod_{m}\frac{1}{1+i\lambda_{m,d}}\right).\label{eq: PS4}\end{equation}

Here $\tilde{\tau}_{SD}$ is the effective spectral diffusion timescale
for the two electron system, $\tilde{\tau}_{SD}^{-4}=\tau_{SD,L}^{-4}+\tau_{SD,R}^{-4}$.
For non-identical dots, 
the eigenvalues $\lambda_{m,d}$ are generally different due to the
differing number of nuclei $N_{d}$, and their associated distributions
of coupling constants $\{A_{k}\}$ and $\{\omega_{k}\}$. All of these
parameters affect the grouping of spins into {}``giant spins,''
and the matrix elements $T_{lk}$, calculated according to Eq.~(\ref{Tkld}).

Equation (\ref{eq: PS4}) is the main result of the paper. This result
shows that, within the semi-classical approach, the echo signal $2P_{S}-1$
in the Hahn echo experiment can be understood in terms of a decay
envelope arising from spectral diffusion, along with an additional
factor which arises from the relative precession of different nuclear
species comprising the transverse Overhauser field. At low magnetic
fields, this precession term leads to the collapse and revival effect.

\subsection{Stability of the revivals peaks against system inhomogeneities}

In this subsection we investigate how modifications of nuclear precession
induced by the small spatial variations of the Knight fields and nuclear
Zeeman couplings affect the electron spin echo signal, Eq.~(\ref{eq: PS4}).
In particular, we focus on the stability of the revival peaks.

Note that a necessary condition for the revivals to appear is the
clear separation of Larmor frequencies, such that $\omega_{kl}\gg\delta\omega_{kl},A_{kl}$
for $\alpha(k)\neq\alpha(l)$. For $\delta\omega_{kl}=A_{kl}=0$,
i.e. for a homogeneous nuclear system with no intrinsic nuclear spin
dephasing, Eq.~(\ref{eq:eiphi3}) reduces to Eq.~(\ref{eq:eiphi2}),
where the right hand side exhibits revivals for values of the free
evolution time $\tau$ satisfying $\omega_{\alpha\beta}\tau\approx4\pi n$,
for any integer n. In this case the overall decay of the spin echo
signal results solely from the spectral diffusion factor $e^{-(\tau/\tilde{\tau}_{SD})^{4}}$
in Eq.~(\ref{eq: PS4}).

Small but nonzero frequency differences $\delta\omega_{kl}$ and $A_{kl}$
can give rise to nuclear spin dephasing and cause additional decay
of the revival peak envelope as a function of $\tau$. We now analyze
this decay and discuss its physical origin. Near the revival peaks
$\omega_{kl}\tau\approx4\pi n$, and for early times $\tau$ satisfying
$\delta\omega_{kl}\tau,A_{kl}\tau\ll1$, Eq.~(\ref{Tkld}) simplifies
to \begin{equation}
T_{kl}\vert_{{\rm peak}}\approx\frac{ig^{*}\mu_{B}\overline{b}_{l}\overline{b}_{k}(\omega_{kl}+\delta\omega_{kl})}{4|B_{{\rm ext}}|}\frac{(A_{kl}\tau)^{2}-(\delta\omega_{kl}\tau)^{2}}{(\omega_{kl}+\delta\omega_{kl})^{2}-A_{kl}^{2}}.\label{eq:Tlk_peak}\end{equation}
 For homo-nuclear terms with $\omega_{kl}$ = 0 (i.e. for groups $k$
and $l$ comprised of the same nuclear species), we find \begin{equation}
T_{kl}\vert_{{\rm peak}}\approx\frac{ig^{*}\mu_{B}\overline{b}_{l}\overline{b}_{k}\delta\omega_{kl}}{4|B_{{\rm ext}}|}\tau^{2},\quad\alpha(k)=\alpha(l),\label{eq:homonuclear}\end{equation}
 while hetero-nuclear terms are given by \begin{equation}
T_{kl}\vert_{{\rm peak}}\approx\frac{ig^{*}\mu_{B}\overline{b}_{l}\overline{b}_{k}}{4|B_{{\rm ext}}|}\frac{A_{kl}^{2}-\delta\omega_{kl}^{2}}{\omega_{kl}}\tau^{2},\ \alpha(k)\neq\alpha(l).\end{equation}

Note that the hetero-nuclear terms and suppressed by the small ratios
of $\frac{\delta\omega_{kl}}{\omega_{kl}}$ or $\frac{A_{kl}}{\omega_{kl}}$.
Hence, for $\delta\omega_{kl}\tau,A_{kl}\tau\ll1$, near the revival
peaks the $T$-matrix is approximately block-diagonal with respect
to the three species. Up to second order in $\omega_{kl}\tau,A_{kl}\tau$,
we find that each block has a single pair of nonzero eigenvalues given
by \begin{equation}
\lambda_{\alpha}\vert_{{\rm peak}}=\pm\frac{g^{*}\mu_{B}a_{\alpha}n_{\alpha}\mathcal{A}_{\alpha}^{2}\sqrt{\langle\delta\omega_{\alpha}^{2}\rangle}}{4N_{d}|B_{{\rm ext}}|}\tau^{2},\end{equation}
 where $\sqrt{\langle\delta\omega_{\alpha}^{2}\rangle}$
is the rms spread of Larmor angular frequencies of species $\alpha$. Thus, we obtain a simple expression for the echo
envelope decay at the revival peaks: 
\begin{equation} 
\langle e^{i\Phi}\rangle\vert_{{\rm peak}}=\prod_{\alpha}\left[1+\left(\frac{g^{*}\mu_{B}a_{\alpha}n_{\alpha}\mathcal{A}_{\alpha}^{2}}{4N_{d}|B_{{\rm ext}}|}\right)^{2}\langle\delta\omega_{\alpha}^{2}\rangle\tau^{4}\right]^{-1}.
\label{eq:ephi}\end{equation}

 Physically, Eq.~(\ref{eq:ephi}) describes decay of the revival
envelope with timescale 
$\tau_\perp^{-4}=\left(\frac{g^{*}\mu_{B}a_{\alpha}n_{\alpha}\mathcal{A}_{\alpha}^{2}}{4N_{d}|B_{{\rm ext}}|}\right)^{2}\langle\delta\omega_{\alpha}^{2}\rangle
$, which arises primarily from the intra-species spread of
the Larmor frequencies, $\sqrt{\langle\delta\omega_{\alpha}^{2}\rangle}$.
The effect of the off-diagonal matrix elements between different species
is negligible. Additionally, the effect of the Knight field
is also negligible, due to the fact that the Knight field reverses
its sign halfway through the evolution when the electron spin is flipped
by the $\pi$-pulse of the spin-echo sequence. 

\subsection{Estimate of quantum corrections to the semi-classical results\label{sub:quantum error}}

The semi-classical treatment is expected to be valid in the limit
of a large number of participating nuclei, $N_{d}$. To better understand
the validity of the approximation for finite $N_{d}$, in this section
we provide a heuristic estimate for the deviation of the semi-classical
expression for the single electron coherence function, $C_{cs}$, Eq.~(\ref{eq: quantum2classical}),
from the quantum expression for $C$, given in Eq.~(\ref{eq:PS3}). We define the quantum error as $C_{cs}-C$. 

We are interested in particular in the quantum error associated with the semi-classical approximation to the dynamics induced by the hyperfine and Zeeman couplings. 
Therefore 
we ignore the nuclear spin-diffusion contribution to the decoherence, which is caused by the dipolar interaction. 
This is done by setting $\tau_{SD}=\infty$ in Eqs.~(\ref{eq:PS3}) and (\ref{eq: quantum2classical}). 
We analyze the scaling of the quantum error as $N_{d}$ is increased. However, while changing $N_d$, we wish to keep fixed the rms values of the Overhauser field components, which determine the timescale for the decoherence of the electrons spins. 
Given Eq.~(\ref{eq:Bnuc}), the
rms value of the Overhauser field for each species scales as $\mathcal{A}_{\alpha}/\sqrt{N_{d}}$,
therefore we require $\mathcal{A}_{\alpha}\propto\sqrt{N_{d}}$.

For large $N_{d}$, the leading contribution to the quantum
error comes from the fact that for a given initial nuclear spin state
$\left|\Psi_{i}\right\rangle =\bigotimes_{k}\left|I_{k},\hat{\vec{n}}_{k}\right\rangle $,
the overlap of the final nuclear spin states for different initial
electron spin states, $\left|\Psi_{f+}\right\rangle \equiv U_{\perp+}\left(\tau\right)\left|\Psi\right\rangle $
for $\sigma=+$ and $\left|\Psi_{f-}\right\rangle \equiv U_{\perp-}\left(\tau\right)\left|\Psi\right\rangle $
for $\sigma=-$, is not unity, $|\Amp{\Psi_{f+}}{\Psi_{f-}}|<1$.
In other words, for every initial coherent state, the final state
includes quantum correlations between the electron and nuclear spins, which are not
captured by the semi-classical treatment.
Thus, the electron-state-dependent modification to the nuclear evolution
contributes an additional suppression of $C$ which is not accounted
for in Eq.~(\ref{eq:semi_classical_approx}).

Up to leading order in the hyperfine coupling, the nuclear
spin evolution is dominated by Larmor precession due to the combination
of the external magnetic field and the component of the Knight field
parallel to the external field axis, $z$. Both contributions are
included in $H_{0}$ in Eq.~(\ref{H0}). However, these two effects
alone will result in a perfect overlap of the final states, $\left|\Psi_{f+}\right\rangle =\left|\Psi_{f-}\right\rangle $.
This is because the Larmor precession about the external field
is independent of the electron spin state, while the precession due
to the Knight field is perfectly reversed halfway through the evolution
due to the echo pulse which flips the electron spin. Thus both effects
were fully accounted for in the semi-classical treatment above.

The quantum error results from the higher order terms in the
hyperfine coupling, in particular from the transverse components of
the Knight field, i.e. from the terms $S_{d}^{+}I_{n}^{-}+S_{d}^{-}I_{n}^{+}$
in the system's Hamiltonian, Eq.~(\ref{eq: H}). These terms contribute
to $\sigma H_{\perp}^{\sigma}(t)=\sigma\frac{g^{*}\mu_{B}|\vec{B}_{{\rm nuc},d}^{\perp\sigma}(t)|^{2}}{4|B_{{\rm ext}}|}$
in the reduced Hamiltonian, Eq.~(\ref{eq:H_el_t}). Although the
transverse components of the Knight field are also reversed after
the echo pulse, the fact that they do not commute with $H_{0}$ means that 
difference in the final states $\left|\Psi_{f+}\right\rangle $ and $\left|\Psi_{f-}\right\rangle $ may survive the echo.

Below we first focus our discussion on a theoretical model which, in the semi-classical treatment, produces perfect revivals in the echo signal due to exact commensuration between Larmor periods of different nuclear species. 
We separate the discussion into two cases, in which the free evolution time $\tau$
is either exactly on, or is away from, a revival peak. 
Then, at the end of the section we consider the quantum error at a revival peak which is not perfect, even within the semi-classical approximation, due to a lack of commensuration between nuclear Larmor periods.

For the initial state  $\left|\Psi_{i}\right\rangle$ given above, which is a product of ``giant spin'' coherent states in each of the $K$ groups of nuclear spins, we argue that,
to leading order in $N_{d}$, the final states  $\left|\Psi_{f\sigma}\right\rangle $ remain
 approximately tensor products of coherent states,
\begin{equation}
 \left|\Psi_{f\sigma}\right\rangle \approx \bigotimes_{k}\left|I_{k},\hat{\vec{n}}_{k,f\sigma}\right\rangle
\end{equation}
Within this picture,  
$\sigma H_{\perp}^{\sigma}(t)$ causes the giant spin coherent states to 
rotate to new directions $\{\hat{\vec{n}}_{k,f\sigma}\}$, which depend on the electron spin state $\sigma=+$ or $-$. 
Other quantum effects such
as coherent spin state squeezing due to the $\left(I_{k}^{x}\right)^{2}+\left(I_{k}^{y}\right)^{2}$
terms in $H_{\perp}^{\sigma}(t)$, which may stretch the coherent
states anisotropically, or a build-up of quantum correlations between
different giant spins, are also expected. However, these effects enter
only at higher order in $1/I_{k}$, because they are seeded by the
small quantum fluctuations of the spin components in the initial coherent
state. 

Using the argument above, we find that away from the revival
peaks the angle between the directions of the two final states of
each giant spin $k$, for \textbf{$\sigma=+$ }or\textbf{ $-$}, scales
as $1/I_{k}$. Why is this so? First, according to Eq.~(\ref{eq: H}),
the Knight field acting on the nuclear spins, $A_{d,n}\mathbf{S}_{d}$,
is smaller than the Overhauser field acting on the electron by a factor
$1/\sqrt{N_{d}}\sim1/I_{k}$. Second, assuming that the timescale
of the electron spin coherence collapse is comparable to the timescale
between revivals, the contribution of $H_{\perp}(t)$ to the Overhauser
field causes electron spin precession through an angle of order $2\pi$.
Over the same time interval, the corresponding part of the Knight
field will cause the nuclear spins to rotate through an angle which
is $1/I_{k}$ times smaller. 

The overlap \textbf{$\left|\Amp{\Psi_{f+}}{\Psi_{f-}}\right|$}
can thus be approximated to leading order by a product of overlaps
between pairs of coherent states $\prod_{k}\left|\Amp{I_{k},\hat{\vec{n}}_{k,f+}}{I_{k},\hat{\vec{n}}_{k,f-}}\right|$,
which are misaligned 
by an angle that scales down as $1/I_{k}$. 
Furthermore, because $I_k$ is large, each coherent state $\left|I_{k},\hat{\vec{n}}_{kf\pm}\right\rangle $
is characterized by a Gaussian phase space distribution with angular width of order $1/\sqrt{I_k}$.
The reduction of the overlap, $1 - \left|\Amp{I_{k},\hat{\vec{n}}_{k,f+}}{I_{k},\hat{\vec{n}}_{k,f-}}\right|$,
therefore scales as $1/I_{k}\sim\sqrt{\frac{K}{N_{d}}}$. We therefore find
that the overall reduction of the total overlap, $1 - \left|\Amp{\Psi_{f+}}{\Psi_{f-}}\right|$,
scales down at least as $\frac{K}{I_{k}}\sim\sqrt{K^{3}/N_{d}}$,
as does the quantum error.

However, exactly at a revival peak, where the condition of perfect
nuclear precession commensurability is met, this leading contribution
vanishes. 
Here, for a given giant spin $k$, treating all other spins
$k'\neq k$ as classical vectors, the precession and squeezing induced
by $H_{\perp}^{\sigma}(t)$ is perfectly reversed after the $\pi$-pulse.
Thus the quantum error at a revival peak is a higher order effect
in $1/\sqrt{N_{d}}$. Numerical simulations with just two giant
spins indicate that the quantum error at the revival peaks may scale
down even faster then $1/N_{d}$, but a more detailed analysis is
a subject for further study.

In more realistic cases, when the commensurability condition
of all nuclear species cannot be exactly met, the quantum error near
the quasi-revival peak scales like $f(\tau)\sqrt{K^{3}/N_{d}}$, with
the pre-factor $f(\tau)$ becoming small as $\tau$ approaches the
approximate commensuration point. 
At the quasi-revival peak, $f(\tau)$
is then dominated by the classical effect of imperfect commensuration,
captured by $1-C_{sc}\left(\tau\right)$. 
We estimate the error 
in the case of GaAs quantum dots with $N_{d}=(2-4)\times10^{6}$.
For this estimate, we take the minimal number of groups of nuclear spins $K$ which, within the semi-classical treatment, accurately produces the echo signal behavior near the revivals peaks, without changing significantly upon further refinement to more groups.
For that purpose, note that 
only the spread  of the Larmor frequencies for each nuclear species, $\langle\delta\omega_{\alpha}^{2}\rangle$, enters the expression for $1-C_{sc}\left(\tau\right)$ in Eq.~(\ref{eq:ephi}). 
Consequently, it is sufficient 
to divide each of the three species into two collective
groups. Grouping the nuclei into $K\approx6$ collective giant spins gives
$\sqrt{K^{3}/N_{d}}\lesssim1/90$. Therefore the quantum contribution
to the imperfect revival is smaller than the semi-classical contribution
by more than an order of magnitude.

\section{Discussion}

\label{sec:Discussion} Equation (\ref{eq: PS4}), together with Eq.~(\ref{Tkld}),
was used to fit the experimental echo signal data in Ref.~\onlinecite{ourselves}.
As stated in Sec.~\ref{sub: calculation}, in general the matrix
$T_{kl}$ and its eigenvalues $\{\lambda_{m}\}$ are different for
each dot due to variations in dot size and local environment. In particular,
the distribution of hyperfine couplings, $\{A_{k}\}$, and the number
of nuclei, $N_{d}$, depend on the distribution of electron density
in the dot. Furthermore, the distribution of $\{\delta\omega_{k}\}$
depends on, e.g. local electric field gradients which couple to the
nuclear quadrupole moments. However, in practice we found that choosing
the same parameters for $T_{kl}$ in the two dots produced a very
good fit to the experimental data. Allowing, for example, different
values of $N_{d}$ for the two dots did not significantly improve
the fits.

The spread of the Larmor frequencies $\sqrt{\langle\delta\omega_{\alpha}^{2}\rangle}$
that was found from fitting to the experimental results \cite{ourselves}
was equivalent to $3$ Gauss effective spread of magnetic field, somewhat
bigger than the values of NMR line-widths, typically about $1$ Gauss\textbf{
}equivalent spread, reported for bulk GaAs in the literature \cite{Hester:defect,Sundfors:NMR}.
In addition to a random contribution to the local magnetic field coming
from the dipolar interaction with neighboring nuclei, the nuclear
spins in a GaAs quantum dot experience a quadrupolar splitting due
to electric field gradients originating from the confined electrons,
which we estimate to be at the order of a few Gauss (see Ref.~\onlinecite{ourselves},
supplementary materials). These quadrupolar shifts may be responsible
for the difference between the observed value and the NMR line-width.
In the echo experiments which have been performed so far there is
no way to distinguish between different origins of the apparent spread
of nuclear Larmor frequencies (i.e. between decay due to nuclear spin
flip-flop, due to inhomogeneous broadening, or due to quadrupolar
effects).

Our model does not directly include the dipole-dipole-induced temporal
decay of \emph{local} nuclear spin correlations $\Avg{I_{n}^{i}(0)I_{n}^{j}(t)}$.
However, such decay is accounted for phenomenologically by a time-independent
spread of site-dependent Larmor precession frequencies. Because the
current Hahn echo experiments can't distinguish between these intrinsic
and extrinsic nuclear spin dephasing processes, our approximation
of accounting for these effects by a random static, disordered, Zeeman
field is reasonable. According to Eq.~(\ref{eq:ephi}), the affect of this random field is to cause an additional decay to the echo signal with a timescale $\tau_\perp$. This decay becomes dominant at low magnetic fields, as  $\tau_\perp$ becomes shorter then the decay time associated with  the spectral diffusion, $\tau_{SD}$.

In addition, note that the spectral diffusion decay time $\tau_{SD}$
was found experimentally to be independent of the magnetic field strength.
Other experiments\cite{Reilly2008A} have also suggested that the
external magnetic field does not significantly influence spin-diffusion
at the relevant field range, above $20$ mT. This provides further
justification for neglecting the commutators $[H_{\perp},H_{d}]$
in section \ref{sec:SC_derivation}, which, \emph{a priori}, could
introduce such a dependence.

In conclusion, we have provided physical justification for treating
the Overhauser fields in a GaAs double quantum dot system as classical
vector variables, based on the relative smallness of the commutation
relations due to the large number of spins in each dot. For
the simple model in section \ref{sec:simplest_model}, in which we ignored the nuclear dipole-dipole interaction and the Knight shift of the nuclear Larmor frequencies, we demonstrated
the equivalence of the semi-classical treatment and the quantum diagrammatic
summation of Ref.~\onlinecite{Cywinski2009}. The semi-classical
treatment quantitatively captures the observed phenomena of monotonic
decay of the echo signal in strong magnetic fields, and the collapses
and revivals of the echo signal in weaker magnetic fields. 
The overall effect of the Knight field on the spin-echo is found to be negligible. 
This fact is parameter dependent. However note that the Knight field is reversed by the echo pulse.  As a result, If one treats the Knight field contribution to the phase in Eqs.~(\ref{eq: Phi_d_zz}) and (\ref{eq:Tlk_peak}) as a perturbation, it vanishes in the leading order in $\tau$, near the revivals peak. This is unlike the contribution of competing process, i.e. the spread of the Larmor frequencies.
This means that, for the typical parameters in GaAs quantum dots,
after averaging over all nuclear spin states, both the collapse-revivals
effect and the overall envelope decay can be understood simply as
arising from averaging over the initial conditions of the Larmor precession
of the nuclear spins.

\section*{Acknowledgments}

We acknowledge M. Gullans, J. J. Krich, E. Rashba, J. Maze and L. Cywinski
for stimulating discussions. This work was supported in part by IARPA and by NSF grants PHY-0646094 and DMR-0906475.  A.Y is partly supported by the NSA.

\bibliography{bibdata}

\begin{thebibliography}{41}
\expandafter\ifx\csname natexlab\endcsname\relax\def\natexlab#1{#1}\fi
\expandafter\ifx\csname bibnamefont\endcsname\relax
  \def\bibnamefont#1{#1}\fi
\expandafter\ifx\csname bibfnamefont\endcsname\relax
  \def\bibfnamefont#1{#1}\fi
\expandafter\ifx\csname citenamefont\endcsname\relax
  \def\citenamefont#1{#1}\fi
\expandafter\ifx\csname url\endcsname\relax
  \def\url#1{\texttt{#1}}\fi
\expandafter\ifx\csname urlprefix\endcsname\relax\def\urlprefix{URL }\fi
\providecommand{\bibinfo}[2]{#2}
\providecommand{\eprint}[2][]{\url{#2}}

\bibitem[{\citenamefont{Petta et~al.}(2005)\citenamefont{Petta, Johnson,
  Taylor, Laird, Yacoby, Lukin, Marcus, Hanson, and Gossard}}]{Petta2005}
\bibinfo{author}{\bibfnamefont{J.~R.} \bibnamefont{Petta}},
  \bibinfo{author}{\bibfnamefont{A.~C.} \bibnamefont{Johnson}},
  \bibinfo{author}{\bibfnamefont{J.~M.} \bibnamefont{Taylor}},
  \bibinfo{author}{\bibfnamefont{E.~A.} \bibnamefont{Laird}},
  \bibinfo{author}{\bibfnamefont{A.}~\bibnamefont{Yacoby}},
  \bibinfo{author}{\bibfnamefont{M.~D.} \bibnamefont{Lukin}},
  \bibinfo{author}{\bibfnamefont{C.~M.} \bibnamefont{Marcus}},
  \bibinfo{author}{\bibfnamefont{M.~P.} \bibnamefont{Hanson}},
  \bibnamefont{and} \bibinfo{author}{\bibfnamefont{A.~C.}
  \bibnamefont{Gossard}}, \bibinfo{journal}{Science}
  \textbf{\bibinfo{volume}{309}}, \bibinfo{pages}{2180} (\bibinfo{year}{2005}).

\bibitem[{\citenamefont{Koppens et~al.}(2008)\citenamefont{Koppens, Nowack, and
  Vandersypen}}]{Koppens2008}
\bibinfo{author}{\bibfnamefont{F.~H.~L.} \bibnamefont{Koppens}},
  \bibinfo{author}{\bibfnamefont{K.~C.} \bibnamefont{Nowack}},
  \bibnamefont{and} \bibinfo{author}{\bibfnamefont{L.~M.~K.}
  \bibnamefont{Vandersypen}}, \bibinfo{journal}{Phys. Rev. Lett.}
  \textbf{\bibinfo{volume}{100}}, \bibinfo{pages}{236802}
  (\bibinfo{year}{2008}).

\bibitem[{\citenamefont{Greilich et~al.}(2006)\citenamefont{Greilich, Yakovlev,
  Shabaev, Efros, Yugova, Oulton, Stavarache, Reuter, Wieck, and
  Bayer}}]{Greilich2006}
\bibinfo{author}{\bibfnamefont{A.}~\bibnamefont{Greilich}},
  \bibinfo{author}{\bibfnamefont{D.~R.} \bibnamefont{Yakovlev}},
  \bibinfo{author}{\bibfnamefont{A.}~\bibnamefont{Shabaev}},
  \bibinfo{author}{\bibfnamefont{A.~L.} \bibnamefont{Efros}},
  \bibinfo{author}{\bibfnamefont{I.~A.} \bibnamefont{Yugova}},
  \bibinfo{author}{\bibfnamefont{R.}~\bibnamefont{Oulton}},
  \bibinfo{author}{\bibfnamefont{V.}~\bibnamefont{Stavarache}},
  \bibinfo{author}{\bibfnamefont{D.}~\bibnamefont{Reuter}},
  \bibinfo{author}{\bibfnamefont{A.~D.} \bibnamefont{Wieck}}, \bibnamefont{and}
  \bibinfo{author}{\bibfnamefont{M.}~\bibnamefont{Bayer}},
  \bibinfo{journal}{Science} \textbf{\bibinfo{volume}{313}},
  \bibinfo{pages}{341} (\bibinfo{year}{2006}).

\bibitem[{\citenamefont{Foletti et~al.}(2009)\citenamefont{Foletti, Bluhm,
  Mahalu, Umansky, and Yacoby}}]{Foletti2009}
\bibinfo{author}{\bibfnamefont{S.}~\bibnamefont{Foletti}},
  \bibinfo{author}{\bibfnamefont{H.}~\bibnamefont{Bluhm}},
  \bibinfo{author}{\bibfnamefont{D.}~\bibnamefont{Mahalu}},
  \bibinfo{author}{\bibfnamefont{V.}~\bibnamefont{Umansky}}, \bibnamefont{and}
  \bibinfo{author}{\bibfnamefont{A.}~\bibnamefont{Yacoby}},
  \bibinfo{journal}{Nature Physics} \textbf{\bibinfo{volume}{5}},
  \bibinfo{pages}{903} (\bibinfo{year}{2009}).

\bibitem[{\citenamefont{Nowack et~al.}(2007)\citenamefont{Nowack, Koppens,
  Nazarov, and Vandersypen}}]{Nowack2007}
\bibinfo{author}{\bibfnamefont{K.~C.} \bibnamefont{Nowack}},
  \bibinfo{author}{\bibfnamefont{F.~H.~L.} \bibnamefont{Koppens}},
  \bibinfo{author}{\bibfnamefont{Y.~V.} \bibnamefont{Nazarov}},
  \bibnamefont{and} \bibinfo{author}{\bibfnamefont{L.~M.~K.}
  \bibnamefont{Vandersypen}}, \bibinfo{journal}{Science}
  \textbf{\bibinfo{volume}{318}}, \bibinfo{pages}{1430} (\bibinfo{year}{2007}).

\bibitem[{\citenamefont{Barthel et~al.}(2009)\citenamefont{Barthel, Reilly,
  Marcus, Hanson, and Gossard}}]{Barthel2009}
\bibinfo{author}{\bibfnamefont{C.}~\bibnamefont{Barthel}},
  \bibinfo{author}{\bibfnamefont{D.~J.} \bibnamefont{Reilly}},
  \bibinfo{author}{\bibfnamefont{C.~M.} \bibnamefont{Marcus}},
  \bibinfo{author}{\bibfnamefont{M.~P.} \bibnamefont{Hanson}},
  \bibnamefont{and} \bibinfo{author}{\bibfnamefont{A.~C.}
  \bibnamefont{Gossard}}, \bibinfo{journal}{Phys. Rev. Lett.}
  \textbf{\bibinfo{volume}{103}}, \bibinfo{eid}{160503} (\bibinfo{year}{2009}).

\bibitem[{\citenamefont{Loss and DiVincenzo}(1998)}]{Loss1998}
\bibinfo{author}{\bibfnamefont{D.}~\bibnamefont{Loss}} \bibnamefont{and}
  \bibinfo{author}{\bibfnamefont{D.~P.} \bibnamefont{DiVincenzo}},
  \bibinfo{journal}{Phys. Rev. A} \textbf{\bibinfo{volume}{57}},
  \bibinfo{pages}{120} (\bibinfo{year}{1998}).

\bibitem[{\citenamefont{Taylor et~al.}(2005)\citenamefont{Taylor, Engel, Dur,
  Yacoby, Marcus, Zoller, and Lukin}}]{Taylor2005}
\bibinfo{author}{\bibfnamefont{J.~M.} \bibnamefont{Taylor}},
  \bibinfo{author}{\bibfnamefont{H.~A.} \bibnamefont{Engel}},
  \bibinfo{author}{\bibfnamefont{W.}~\bibnamefont{Dur}},
  \bibinfo{author}{\bibfnamefont{A.}~\bibnamefont{Yacoby}},
  \bibinfo{author}{\bibfnamefont{C.~M.} \bibnamefont{Marcus}},
  \bibinfo{author}{\bibfnamefont{P.}~\bibnamefont{Zoller}}, \bibnamefont{and}
  \bibinfo{author}{\bibfnamefont{M.~D.} \bibnamefont{Lukin}},
  \bibinfo{journal}{Nature Physics} \textbf{\bibinfo{volume}{1}},
  \bibinfo{pages}{177} (\bibinfo{year}{2005}).

\bibitem[{\citenamefont{Taylor et~al.}(2007)\citenamefont{Taylor, Petta,
  Johnson, Yacoby, Marcus, and Lukin}}]{Taylor2007}
\bibinfo{author}{\bibfnamefont{J.~M.} \bibnamefont{Taylor}},
  \bibinfo{author}{\bibfnamefont{J.~R.} \bibnamefont{Petta}},
  \bibinfo{author}{\bibfnamefont{A.~C.} \bibnamefont{Johnson}},
  \bibinfo{author}{\bibfnamefont{A.}~\bibnamefont{Yacoby}},
  \bibinfo{author}{\bibfnamefont{C.~M.} \bibnamefont{Marcus}},
  \bibnamefont{and} \bibinfo{author}{\bibfnamefont{M.~D.} \bibnamefont{Lukin}},
  \bibinfo{journal}{Phys.\ Rev.\ B.} \textbf{\bibinfo{volume}{76}},
  \bibinfo{pages}{035315} (\bibinfo{year}{2007}).

\bibitem[{\citenamefont{Hanson et~al.}(2007)\citenamefont{Hanson, Kouwenhoven,
  Petta, Tarucha, and Vandersypen}}]{HansonRMP}
\bibinfo{author}{\bibfnamefont{R.}~\bibnamefont{Hanson}},
  \bibinfo{author}{\bibfnamefont{L.~P.} \bibnamefont{Kouwenhoven}},
  \bibinfo{author}{\bibfnamefont{J.~R.} \bibnamefont{Petta}},
  \bibinfo{author}{\bibfnamefont{S.}~\bibnamefont{Tarucha}}, \bibnamefont{and}
  \bibinfo{author}{\bibfnamefont{L.~M.~K.} \bibnamefont{Vandersypen}},
  \bibinfo{journal}{Rev. Mod. Phys.} \textbf{\bibinfo{volume}{79}},
  \bibinfo{pages}{1217} (\bibinfo{year}{2007}).

\bibitem[{\citenamefont{Merkulov et~al.}(2002)\citenamefont{Merkulov, Efros,
  and M.Rosen}}]{Merkulov2002}
\bibinfo{author}{\bibfnamefont{I.~A.} \bibnamefont{Merkulov}},
  \bibinfo{author}{\bibfnamefont{A.~L.} \bibnamefont{Efros}}, \bibnamefont{and}
  \bibinfo{author}{\bibnamefont{M.Rosen}}, \bibinfo{journal}{Phys. Rev. B}
  \textbf{\bibinfo{volume}{65}}, \bibinfo{pages}{205309}
  (\bibinfo{year}{2002}).

\bibitem[{\citenamefont{Erlingsson and Nazarov}(2002)}]{Erlingsson2002}
\bibinfo{author}{\bibfnamefont{S.~I.} \bibnamefont{Erlingsson}}
  \bibnamefont{and} \bibinfo{author}{\bibfnamefont{Y.~V.}
  \bibnamefont{Nazarov}}, \bibinfo{journal}{Phys. Rev. B}
  \textbf{\bibinfo{volume}{66}}, \bibinfo{pages}{155327}
  (\bibinfo{year}{2002}).

\bibitem[{\citenamefont{{Johnson} et~al.}(2005)\citenamefont{{Johnson},
  {Petta}, {Taylor}, {Yacoby}, {Lukin}, {Marcus}, {Hanson}, and
  {Gossard}}}]{Johnson2005}
\bibinfo{author}{\bibfnamefont{A.~C.} \bibnamefont{{Johnson}}},
  \bibinfo{author}{\bibfnamefont{J.~R.} \bibnamefont{{Petta}}},
  \bibinfo{author}{\bibfnamefont{J.~M.} \bibnamefont{{Taylor}}},
  \bibinfo{author}{\bibfnamefont{A.}~\bibnamefont{{Yacoby}}},
  \bibinfo{author}{\bibfnamefont{M.~D.} \bibnamefont{{Lukin}}},
  \bibinfo{author}{\bibfnamefont{C.~M.} \bibnamefont{{Marcus}}},
  \bibinfo{author}{\bibfnamefont{M.~P.} \bibnamefont{{Hanson}}},
  \bibnamefont{and} \bibinfo{author}{\bibfnamefont{A.~C.}
  \bibnamefont{{Gossard}}}, \bibinfo{journal}{Nature}
  \textbf{\bibinfo{volume}{435}}, \bibinfo{pages}{925} (\bibinfo{year}{2005}).

\bibitem[{\citenamefont{de~Sousa and DasSarma}(2003)}]{Sousa2003}
\bibinfo{author}{\bibfnamefont{R.}~\bibnamefont{de~Sousa}} \bibnamefont{and}
  \bibinfo{author}{\bibfnamefont{S.}~\bibnamefont{DasSarma}},
  \bibinfo{journal}{Phys.\ Rev.\ B.} \textbf{\bibinfo{volume}{68}},
  \bibinfo{pages}{115322} (\bibinfo{year}{2003}).

\bibitem[{\citenamefont{{Koppens} et~al.}(2005)\citenamefont{{Koppens}, {Folk},
  {Elzerman}, {Hanson}, {Willems van Beveren}, {Vink}, {Tranitz},
  {Wegschneider}, {Kouwenhoven}, and {Vandersypen}}}]{Koppens2005}
\bibinfo{author}{\bibfnamefont{F.~H.~L.} \bibnamefont{{Koppens}}},
  \bibinfo{author}{\bibfnamefont{J.~A.} \bibnamefont{{Folk}}},
  \bibinfo{author}{\bibfnamefont{J.~M.} \bibnamefont{{Elzerman}}},
  \bibinfo{author}{\bibfnamefont{R.}~\bibnamefont{{Hanson}}},
  \bibinfo{author}{\bibfnamefont{L.~H.} \bibnamefont{{Willems van Beveren}}},
  \bibinfo{author}{\bibfnamefont{I.~T.} \bibnamefont{{Vink}}},
  \bibinfo{author}{\bibfnamefont{H.~P.} \bibnamefont{{Tranitz}}},
  \bibinfo{author}{\bibfnamefont{W.}~\bibnamefont{{Wegschneider}}},
  \bibinfo{author}{\bibfnamefont{L.~P.} \bibnamefont{{Kouwenhoven}}},
  \bibnamefont{and} \bibinfo{author}{\bibfnamefont{L.~M.~K.}
  \bibnamefont{{Vandersypen}}}, \bibinfo{journal}{Science}
  \textbf{\bibinfo{volume}{309}}, \bibinfo{pages}{1346} (\bibinfo{year}{2005}).

\bibitem[{\citenamefont{Hahn}(1950)}]{Hahn1950}
\bibinfo{author}{\bibfnamefont{E.~L.} \bibnamefont{Hahn}},
  \bibinfo{journal}{Phys. Rev.} \textbf{\bibinfo{volume}{80}},
  \bibinfo{pages}{580} (\bibinfo{year}{1950}).

\bibitem[{\citenamefont{Meiboom and Gill}(1958)}]{Meiboom1958}
\bibinfo{author}{\bibfnamefont{S.}~\bibnamefont{Meiboom}} \bibnamefont{and}
  \bibinfo{author}{\bibfnamefont{D.}~\bibnamefont{Gill}},
  \bibinfo{journal}{Rev. Sci. Inst.} \textbf{\bibinfo{volume}{29}},
  \bibinfo{pages}{688} (\bibinfo{year}{1958}).

\bibitem[{\citenamefont{Bluhm et~al.}(2011)\citenamefont{Bluhm, Foletti, Neder,
  Rudner, Mahalu, Umansky, and Yacoby}}]{ourselves}
\bibinfo{author}{\bibfnamefont{H.}~\bibnamefont{Bluhm}},
  \bibinfo{author}{\bibfnamefont{S.}~\bibnamefont{Foletti}},
  \bibinfo{author}{\bibfnamefont{I.}~\bibnamefont{Neder}},
  \bibinfo{author}{\bibfnamefont{M.}~\bibnamefont{Rudner}},
  \bibinfo{author}{\bibfnamefont{D.}~\bibnamefont{Mahalu}},
  \bibinfo{author}{\bibfnamefont{V.}~\bibnamefont{Umansky}}, \bibnamefont{and}
  \bibinfo{author}{\bibfnamefont{A.}~\bibnamefont{Yacoby}},
  \bibinfo{journal}{Nature Physics} \textbf{\bibinfo{volume}{7}},
  \bibinfo{pages}{109} (\bibinfo{year}{2011}).

\bibitem[{\citenamefont{Gaudin}(1976)}]{Gaudin1976}
\bibinfo{author}{\bibfnamefont{M.}~\bibnamefont{Gaudin}}, \bibinfo{journal}{J.
  de Physique} \textbf{\bibinfo{volume}{37}}, \bibinfo{pages}{1087}
  (\bibinfo{year}{1976}).

\bibitem[{\citenamefont{Khaetskii et~al.}(2002)\citenamefont{Khaetskii, Loss,
  and Glazman}}]{Khaetskii2002}
\bibinfo{author}{\bibfnamefont{A.~V.} \bibnamefont{Khaetskii}},
  \bibinfo{author}{\bibfnamefont{D.}~\bibnamefont{Loss}}, \bibnamefont{and}
  \bibinfo{author}{\bibfnamefont{L.}~\bibnamefont{Glazman}},
  \bibinfo{journal}{Phys.\ Rev.\ Lett.} \textbf{\bibinfo{volume}{88}},
  \bibinfo{pages}{186802} (\bibinfo{year}{2002}).

\bibitem[{\citenamefont{Erlingsson and Nazarov}(2004)}]{Erlingsson2004}
\bibinfo{author}{\bibfnamefont{S.~I.} \bibnamefont{Erlingsson}}
  \bibnamefont{and} \bibinfo{author}{\bibfnamefont{Y.~V.}
  \bibnamefont{Nazarov}}, \bibinfo{journal}{Phys. Rev. B}
  \textbf{\bibinfo{volume}{70}}, \bibinfo{pages}{205327}
  (\bibinfo{year}{2004}).

\bibitem[{\citenamefont{Yao et~al.}(2006)\citenamefont{Yao, Liu, and
  Sham}}]{Yao2006}
\bibinfo{author}{\bibfnamefont{W.}~\bibnamefont{Yao}},
  \bibinfo{author}{\bibfnamefont{R.~B.} \bibnamefont{Liu}}, \bibnamefont{and}
  \bibinfo{author}{\bibfnamefont{L.~J.} \bibnamefont{Sham}},
  \bibinfo{journal}{Phys.\ Rev.\ B.} \textbf{\bibinfo{volume}{74}},
  \bibinfo{pages}{195301} (\bibinfo{year}{2006}).

\bibitem[{\citenamefont{Yao et~al.}(2007)\citenamefont{Yao, Liu, and
  Sham}}]{Yao2007}
\bibinfo{author}{\bibfnamefont{W.}~\bibnamefont{Yao}},
  \bibinfo{author}{\bibfnamefont{R.~B.} \bibnamefont{Liu}}, \bibnamefont{and}
  \bibinfo{author}{\bibfnamefont{L.~J.} \bibnamefont{Sham}},
  \bibinfo{journal}{Phys. Rev. Lett.} \textbf{\bibinfo{volume}{98}},
  \bibinfo{pages}{077602} (\bibinfo{year}{2007}).

\bibitem[{\citenamefont{Deng and Hu}(2006)}]{Deng2006}
\bibinfo{author}{\bibfnamefont{C.}~\bibnamefont{Deng}} \bibnamefont{and}
  \bibinfo{author}{\bibfnamefont{X.}~\bibnamefont{Hu}}, \bibinfo{journal}{Phys.
  Rev. B} \textbf{\bibinfo{volume}{73}}, \bibinfo{pages}{241303}
  (\bibinfo{year}{2006}).

\bibitem[{\citenamefont{Coish and Loss}(2004)}]{Coish2004}
\bibinfo{author}{\bibfnamefont{W.~A.} \bibnamefont{Coish}} \bibnamefont{and}
  \bibinfo{author}{\bibfnamefont{D.}~\bibnamefont{Loss}},
  \bibinfo{journal}{Phys. Rev. B} \textbf{\bibinfo{volume}{70}},
  \bibinfo{pages}{195340} (\bibinfo{year}{2004}).

\bibitem[{\citenamefont{Witzel et~al.}(2005)\citenamefont{Witzel, de~Sousa, and
  Das~Sarma}}]{Witzel2005}
\bibinfo{author}{\bibfnamefont{W.~M.} \bibnamefont{Witzel}},
  \bibinfo{author}{\bibfnamefont{R.}~\bibnamefont{de~Sousa}}, \bibnamefont{and}
  \bibinfo{author}{\bibfnamefont{S.}~\bibnamefont{Das~Sarma}},
  \bibinfo{journal}{Phys. Rev. B} \textbf{\bibinfo{volume}{72}},
  \bibinfo{pages}{161306} (\bibinfo{year}{2005}).

\bibitem[{\citenamefont{Shenvi et~al.}(2005)\citenamefont{Shenvi, de~Sousa, and
  Whaley}}]{Shenvi2005}
\bibinfo{author}{\bibfnamefont{N.}~\bibnamefont{Shenvi}},
  \bibinfo{author}{\bibfnamefont{R.}~\bibnamefont{de~Sousa}}, \bibnamefont{and}
  \bibinfo{author}{\bibfnamefont{K.~B.} \bibnamefont{Whaley}},
  \bibinfo{journal}{Phys. Rev. B} \textbf{\bibinfo{volume}{71}},
  \bibinfo{pages}{224411} (\bibinfo{year}{2005}).

\bibitem[{\citenamefont{Al-Hassanieh et~al.}(2006)\citenamefont{Al-Hassanieh,
  Dobrovitski, Dagotto, and Harmon}}]{Al-Hassanieh2006}
\bibinfo{author}{\bibfnamefont{K.~A.} \bibnamefont{Al-Hassanieh}},
  \bibinfo{author}{\bibfnamefont{V.~V.} \bibnamefont{Dobrovitski}},
  \bibinfo{author}{\bibfnamefont{E.}~\bibnamefont{Dagotto}}, \bibnamefont{and}
  \bibinfo{author}{\bibfnamefont{B.~N.} \bibnamefont{Harmon}},
  \bibinfo{journal}{Phys. Rev. Lett.} \textbf{\bibinfo{volume}{97}},
  \bibinfo{pages}{037204} (\bibinfo{year}{2006}).

\bibitem[{\citenamefont{Coish et~al.}(2008)\citenamefont{Coish, Fischer, and
  Loss}}]{Coish2008}
\bibinfo{author}{\bibfnamefont{W.~A.} \bibnamefont{Coish}},
  \bibinfo{author}{\bibfnamefont{J.}~\bibnamefont{Fischer}}, \bibnamefont{and}
  \bibinfo{author}{\bibfnamefont{D.}~\bibnamefont{Loss}},
  \bibinfo{journal}{Phys.\ Rev.\ B.} \textbf{\bibinfo{volume}{77}},
  \bibinfo{pages}{125329} (\bibinfo{year}{2008}).

\bibitem[{\citenamefont{Chen et~al.}(2007)\citenamefont{Chen, Bergman, and
  Balents}}]{Chen:semiclass}
\bibinfo{author}{\bibfnamefont{G.}~\bibnamefont{Chen}},
  \bibinfo{author}{\bibfnamefont{D.~L.} \bibnamefont{Bergman}},
  \bibnamefont{and} \bibinfo{author}{\bibfnamefont{L.}~\bibnamefont{Balents}},
  \bibinfo{journal}{Phys. Rev. B} \textbf{\bibinfo{volume}{76}},
  \bibinfo{pages}{045312} (\bibinfo{year}{2007}).

\bibitem[{\citenamefont{Rashba}(2008)}]{Rashba2008}
\bibinfo{author}{\bibfnamefont{E.~I.} \bibnamefont{Rashba}},
  \bibinfo{journal}{Phys. Rev. B} \textbf{\bibinfo{volume}{78}},
  \bibinfo{pages}{195302} (\bibinfo{year}{2008}).

\bibitem[{\citenamefont{Cywinski
  et~al.}(2009{\natexlab{a}})\citenamefont{Cywinski, Witzel, and {Das
  Sarma}}}]{Cywinski2009}
\bibinfo{author}{\bibfnamefont{L.}~\bibnamefont{Cywinski}},
  \bibinfo{author}{\bibfnamefont{W.~M.} \bibnamefont{Witzel}},
  \bibnamefont{and} \bibinfo{author}{\bibfnamefont{S.}~\bibnamefont{{Das
  Sarma}}}, \bibinfo{journal}{Phys. Rev. B} \textbf{\bibinfo{volume}{79}},
  \bibinfo{pages}{245314} (\bibinfo{year}{2009}{\natexlab{a}}).

\bibitem[{\citenamefont{Cywinski
  et~al.}(2009{\natexlab{b}})\citenamefont{Cywinski, Witzel, and {Das
  Sarma}}}]{Cywinski2009B}
\bibinfo{author}{\bibfnamefont{L.}~\bibnamefont{Cywinski}},
  \bibinfo{author}{\bibfnamefont{W.~M.} \bibnamefont{Witzel}},
  \bibnamefont{and} \bibinfo{author}{\bibfnamefont{S.}~\bibnamefont{{Das
  Sarma}}}, \bibinfo{journal}{Phys. Rev. Lett.} \textbf{\bibinfo{volume}{102}},
  \bibinfo{pages}{057601} (\bibinfo{year}{2009}{\natexlab{b}}).

\bibitem[{\citenamefont{Klauder and Anderson}(1962)}]{Klauder1962}
\bibinfo{author}{\bibfnamefont{J.~R.} \bibnamefont{Klauder}} \bibnamefont{and}
  \bibinfo{author}{\bibfnamefont{P.~W.} \bibnamefont{Anderson}},
  \bibinfo{journal}{Phys. Rev.} \textbf{\bibinfo{volume}{125}},
  \bibinfo{pages}{912} (\bibinfo{year}{1962}).

\bibitem[{\citenamefont{Witzel and DasSarma}(2006)}]{Witzel2006}
\bibinfo{author}{\bibfnamefont{W.~M.} \bibnamefont{Witzel}} \bibnamefont{and}
  \bibinfo{author}{\bibfnamefont{S.}~\bibnamefont{DasSarma}},
  \bibinfo{journal}{Phys.\ Rev.\ B.} \textbf{\bibinfo{volume}{74}},
  \bibinfo{pages}{035322} (\bibinfo{year}{2006}).

\bibitem[{\citenamefont{Witzel and DasSarma}(2008)}]{Witzel2008}
\bibinfo{author}{\bibfnamefont{W.~M.} \bibnamefont{Witzel}} \bibnamefont{and}
  \bibinfo{author}{\bibfnamefont{S.}~\bibnamefont{DasSarma}},
  \bibinfo{journal}{Phys. Rev. B} \textbf{\bibinfo{volume}{77}},
  \bibinfo{eid}{165319} (\bibinfo{year}{2008}).

\bibitem[{\citenamefont{de~Sousa}(2009)}]{Sousa2009}
\bibinfo{author}{\bibfnamefont{R.}~\bibnamefont{de~Sousa}},
  \bibinfo{journal}{Top.\ Appl.\ Phys.} \textbf{\bibinfo{volume}{115}},
  \bibinfo{pages}{183} (\bibinfo{year}{2009}).

\bibitem[{\citenamefont{Mikhailov}(1977)}]{Mikhailov1977}
\bibinfo{author}{\bibfnamefont{V.~V.} \bibnamefont{Mikhailov}},
  \bibinfo{journal}{J. Phys. A: Math. Gen.} \textbf{\bibinfo{volume}{10}},
  \bibinfo{pages}{147} (\bibinfo{year}{1977}).

\bibitem[{\citenamefont{Hester et~al.}(1974)\citenamefont{Hester, Sher, Soest,
  and Weisz}}]{Hester:defect}
\bibinfo{author}{\bibfnamefont{R.~K.} \bibnamefont{Hester}},
  \bibinfo{author}{\bibfnamefont{A.}~\bibnamefont{Sher}},
  \bibinfo{author}{\bibfnamefont{J.~F.} \bibnamefont{Soest}}, \bibnamefont{and}
  \bibinfo{author}{\bibfnamefont{G.}~\bibnamefont{Weisz}},
  \bibinfo{journal}{Phys. Rev. B} \textbf{\bibinfo{volume}{10}},
  \bibinfo{pages}{4262} (\bibinfo{year}{1974}).

\bibitem[{\citenamefont{Sundfors}(1969)}]{Sundfors:NMR}
\bibinfo{author}{\bibfnamefont{R.~K.} \bibnamefont{Sundfors}},
  \bibinfo{journal}{Phys. Rev.} \textbf{\bibinfo{volume}{185}},
  \bibinfo{pages}{458} (\bibinfo{year}{1969}).

\bibitem[{\citenamefont{Reilly et~al.}(2008)\citenamefont{Reilly, Taylor,
  Laird, Petta, Marcus, Hanson, and Gossard}}]{Reilly2008A}
\bibinfo{author}{\bibfnamefont{D.~J.} \bibnamefont{Reilly}},
  \bibinfo{author}{\bibfnamefont{J.~M.} \bibnamefont{Taylor}},
  \bibinfo{author}{\bibfnamefont{E.~A.} \bibnamefont{Laird}},
  \bibinfo{author}{\bibfnamefont{J.~R.} \bibnamefont{Petta}},
  \bibinfo{author}{\bibfnamefont{C.~M.} \bibnamefont{Marcus}},
  \bibinfo{author}{\bibfnamefont{M.~P.} \bibnamefont{Hanson}},
  \bibnamefont{and} \bibinfo{author}{\bibfnamefont{A.~C.}
  \bibnamefont{Gossard}}, \bibinfo{journal}{Phys.\ Rev.\ Lett.}
  \textbf{\bibinfo{volume}{101}}, \bibinfo{eid}{236803} (\bibinfo{year}{2008}).

\end{thebibliography}

\end{document}